%% file: main.tex
\documentclass[lettersize,journal]{IEEEtran}
\usepackage{amsmath,amsfonts}
\usepackage{algorithmic}
\usepackage{array}
\usepackage{subfig}
\usepackage{textcomp}
\usepackage{stfloats}
\usepackage{url}
\usepackage{verbatim}
\usepackage{graphicx}
\def\BibTeX{{\rm B\kern-.05em{\sc i\kern-.025em b}\kern-.08em
    T\kern-.1667em\lower.7ex\hbox{E}\kern-.125emX}}
\usepackage{balance}

\usepackage{amsmath}
\usepackage{amssymb}
\usepackage{color}
\usepackage{xspace}

\usepackage{multirow}
\usepackage{tabularx}
\usepackage{tablefootnote}
\usepackage{booktabs}
\usepackage{makecell}
\usepackage{bbding}

\usepackage{amssymb}
\usepackage{pifont}

\usepackage{microtype}
\usepackage{graphicx}
\usepackage{booktabs}
\usepackage{hyperref}
\usepackage{enumitem}

\newcommand{\todo}[1]{\textcolor{black}{#1}}

\newcommand{\majorrevision}[1]{\textcolor{black}{#1}}
\newenvironment{majorrevisiontable}
  {\color{black}}
{}

\newcommand{\tool}{\textsc{AnyEnhance}\xspace}

\begin{document}
\title{\tool: A Unified Generative Model with Prompt-Guidance and Self-Critic for Voice Enhancement}
\author{Junan Zhang,~\IEEEmembership{}
        Jing Yang,~\IEEEmembership{}
        Zihao Fang,~\IEEEmembership{}
        Yuancheng Wang,~\IEEEmembership{}
        Zehua Zhang,~\IEEEmembership{} \\
        Zhuo Wang,~\IEEEmembership{}
        Fan Fan,~\IEEEmembership{}
        Zhizheng Wu~\IEEEmembership{}

\thanks{J. Zhang, Z. Fang, Y. Wang, Z. Zhang and Z. Wu are affiliated with CUHK-Shenzhen and SRIBD. e-mail: \href{mailto:junanzhang@link.cuhk.edu.cn}{junanzhang@link.cuhk.edu.cn}}
\thanks{J. Yang, Z. Wang and F. Fan are affiliated with Huawei Technologies Co., Ltd.}
\thanks{Z. Wu is the corresponding author. e-mail: \href{mailto:wuzhizheng@cuhk.edu.cn}{wuzhizheng@cuhk.edu.cn}}
}

\maketitle

\begin{abstract}
\input{src/ch00-abstract}
\end{abstract}



\input{src/ch01-introduction}
\input{src/ch02-relatedwork}
\input{src/ch03-method}
\input{src/ch04-experiments}
\input{src/ch05-conclusion}

\section*{Acknowledgments}
This work is funded by the NSFC under Grant 62376237, Shenzhen Science and Technology Program ZDSYS20230626091302006, 2023 Shenzhen stability Science Program, and Internal Project Fund from Shenzhen 
Research Institute of Big Data (Grant No. T00120230002).

\bibliographystyle{IEEEtran}
\bibliography{ref}



\end{document}

%% file: src/ch00-abstract.tex
We introduce \tool, a unified generative model for voice enhancement that processes both speech and singing voices. Based on a masked generative model, \tool is capable of handling both speech and singing voices, supporting a wide range of enhancement tasks including denoising, dereverberation, declipping, super-resolution, and target speaker extraction, all simultaneously and without fine-tuning. \tool introduces a prompt-guidance mechanism for in-context learning, which allows the model to natively accept a reference speaker's timbre. In this way, it could boost enhancement performance when a reference audio is available and enable the target speaker extraction task without altering the underlying architecture. Moreover, we also introduce a self-critic mechanism into the generative process for masked generative models, yielding higher-quality outputs through iterative self-assessment and refinement. Extensive experiments on various enhancement tasks demonstrate \tool outperforms existing methods in terms of both objective metrics and subjective listening tests. Demo audios are publicly available at~\footnote{\url{https://amphionspace.github.io/anyenhance}}. An open-source implementation is provided at~\footnote{\url{https://github.com/viewfinder-annn/anyenhance-v1-ccf-aatc}}.

%% file: src/ch01-introduction.tex

\section{Introduction}

\begin{table*}[htbp]
    \scriptsize
    \centering
    \caption{Comparison of existing versatile enhancement-related models. Denoise: Denoising; Dereverb: Dereverberation; Declip: Declipping; SR: Super Resolution; TSE: Target Speaker Extraction. \textbf{\textit{Simultaneity}} means processing multiple distortions simultaneously.}
    \label{tab:related-work}
    \vspace{-5pt}
    \begin{tabular}{
        >{\centering\arraybackslash}m{1.5cm}
        >{\centering\arraybackslash}m{1.4cm}
        ccccc 
        >{\centering\arraybackslash}m{1.2cm} 
        >{\centering\arraybackslash}m{1.2cm} 
        >{\centering\arraybackslash}m{1.4cm}
        >{\centering\arraybackslash}m{1.8cm}
        >{\centering\arraybackslash}m{1.4cm}}
        \toprule
        \multirow{2}{*}{\textbf{Category}} & \multirow{2}{*}{\textbf{Model}} & \multicolumn{5}{c}{\textbf{Enhancement Tasks}} & \multirow{2}{*}{\textbf{Simultaneity}} & \multirow{2}{*}{\textbf{Domain}} & \multirow{2}{*}{\textbf{Sampling Rate}} & \multirow{2}{*}{\makecell{\textbf{Supports} \\ \textbf{Speaker Prompt?}}} & \multirow{2}{*}{\makecell{\textbf{Fine-tuning} \\ \textbf{Required?}}} \\
        \cline{3-7}
        & & Denoise & Dereverb & Declip & SR & TSE & & & & & \\
        \midrule
        \multirow{6}{*}{Multi-Task}
        & Voicefixer\cite{liu2022voicefixer} & \ding{51} & \ding{51} & \ding{51} & \ding{51} & & \ding{51} & Speech & 44.1kHz & \ding{55} & No \\
        & MaskSR\cite{li2024masksr} & \ding{51} & \ding{51} & \ding{51} & \ding{51} & & \ding{51} & Speech & 44.1kHz & \ding{55} & No \\
        & SpeechX\cite{wang2024speechx} & \ding{51} & & & & \ding{51} & & Speech & 16kHz & \ding{55} & No \\
        & SpeechFlow\cite{liugenerative} & \ding{51} & & & & \ding{51} & \ding{51} & Speech & 16kHz & \ding{55} & Yes \\
        & NeMo\cite{ku2024generative} & \ding{51} & & & & \ding{51} & & Speech & 16kHz & \ding{55} & Yes \\
        & Uniaudio\cite{Yang2024UniAudio} & \ding{51} & \ding{51} & & & \ding{51} & & Speech & 16kHz & \ding{55} & Yes \\
        \midrule
        \multirow{1}{*}{Multi-Domain}
        & AudioSR\cite{liu2024audiosr} & & & & \ding{51} & & & Speech, Singing & 48kHz & \ding{55} & No \\
        \midrule
        Multi-Task Multi-Domain & \tool & \ding{51} & \ding{51} & \ding{51} & \ding{51} & \ding{51} & \ding{51} & Speech, Singing & 44.1kHz & \ding{51} & No \\
        \bottomrule
    \end{tabular}
\end{table*}

Voice enhancement is a fundamental task in robust voice processing, including both speech and singing voice (i.e. vocal) processing.  There are many sub-tasks such as speech denoising, dereverberation, declipping, super-resolution, and target speaker extraction (TSE). They process noise, reverberation, bandwidth limitations, clipping, or other voices, respectively. Note that although speech and singing belong to the same human vocal domain, singing differs from speech in two key respects. First, singing generally exhibits a higher pitch, broader frequency range, and specific rhythm. Second, singing occurs in more complex real-world acoustic environments—such as concerts, livehouses, and bars—leading to challenges like heavier reverberation and varied noise sources. In this work, we research on voice enhancement for both speech and singing voice.

Deep-learning-based methods have been widely used in many enhancement tasks, such as denoising~\cite{defossez2019demucs}, dereverb~\cite{richter2023sgmse}, super-resolution~\cite{liu2024audiosr} and target speaker extraction~\cite{wang2024wesep}, and these models are well applicated into downstream applications, such as automatic speech recognition (ASR)~\cite{li2021espnet,donahue2018exploring}, hearing aids~\cite{cox2023overview}, and in-the-wild data processing~\cite{he2024emilia}. Despite these advancements, most deep-learning-based methods focus on a specific enhancement task, offering limited versatility. In recent studies, versatile enhancement models have been proposed to address multiple tasks and domains with a single model (see Table~\ref{tab:related-work}).

There are several models that aim to handle multiple enhancement tasks. Voicefixer~\cite{liu2022voicefixer} proposed the concept of General Speech Restoration (GSR), which attempts to remove multiple distortions simultaneously (noise, reverberation, clipping, and bandwidth limitation) from the input speech signal. MaskSR~\cite{li2024masksr} is another GSR model using a masked generative model to handle noise, reverberation, clipping, and bandwidth limitation simultaneously. Although they can remove background noise from speech, they can't separate speakers. SpeechX~\cite{wang2024speechx}, SpeechFlow~\cite{liugenerative}, NeMo~\cite{ku2024generative} and UniAudio~\cite{Yang2024UniAudio} are other multi-task enhancement models. SpeechX is a versatile speech generation model capable of tasks such as speech denoising and target speaker extraction. SpeechFlow and NeMo are both flow-matching-based~\cite{lipman2022flow} models that are pretrained on a large-scale dataset and can be fine-tuned for downstream tasks like speech denoising and target speaker extraction. UniAudio is an audio foundation model that supports 11 audio generation tasks covering speech denoising, dereverberation and target speaker extraction. \majorrevision{Although those approaches can handle more speech enhancement tasks including target speaker extraction in one model, task-specific fine-tuning is still needed and they can't handle multiple distortions in one pass (e.g., performing target speaker extraction tasks under noisy conditions). Also, they only work for speech enhancement, ignoring singing voice processing.} Some works have attempted to address multi-domain enhancement problems. AudioSR~\cite{liu2024audiosr} is a latent diffusion model which supports super-resolution across all audio domains (speech, sound effects, and music), \majorrevision{but it only supports super-resolution tasks and does not include other enhancement tasks.} Based on the above analysis, existing models have limitations in terms of both task and domain universality, and some require task-specific fine-tuning. 

To overcome the limitations, we propose a unified framework, \tool. Based on the masked generative model, \tool simultaneously addresses various enhancement tasks shown in Table~\ref{tab:related-work} across both speech and singing voice domains, all without fine-tuning. To deal with real-world singing voice domain, we improved the data simulation process to create more realistic recording scenarios. Moreover, unlike implementing an auxiliary speaker encoder in many target speaker extraction models~\cite{wang2024wesep, tang2024tselm}, we propose a prompt-guidance mechanism in \tool, which provides the model with an in-context-learning capability. This mechanism enables the model to seamlessly accept a reference speaker's voice for target speaker extraction and also improves performance on other tasks, all without architectural modifications. Furthermore, to address the issue of sampling instability inherent in masked generative models~\cite{lezama2022improved}, we integrate a self-critic mechanism, enhancing accuracy and robustness when sampling from logits. These advancements make \tool a versatile solution for both speech and singing voice enhancement tasks, as well as a promising candidate for real-world applications across diverse domains. Experimental results demonstrate that \tool outperforms existing models across various enhancement tasks in terms of both objective metrics and subjective evaluations. Ablation studies further validate the effectiveness of the prompt-guidance mechanism across various tasks, the self-critic sampling strategy, the improved data simulation process, and the integration of various enhancement tasks into a unified model. \majorrevision{Our unified multitask training consistently outperforms single-task enhancement models across all tasks, with average improvements of +0.067 OVRL and +0.13 NISQA, demonstrating strong generalization across diverse speech and singing tasks.} To summarize, our contributions are as follows:

\begin{itemize}
    \item We propose, \tool, a unified \textit{multi-task and multi-domain} enhancement model. Built upon the masked generative model, this model is capable of handling both speech and singing voice domains, supporting a range of enhancement tasks such as denoising, dereverberation, declipping, super-resolution, and target speaker extraction simultaneously without fine-tuning.
    \item We introduce a \textit{prompt-guidance} mechanism in \tool. This prompt-guidance solution enables the model to accept a reference speaker's voice for target speaker extraction and also improves the performance of other enhancement tasks.
    \item We propose a \textit{self-critic sampling} strategy in \tool. The self-critic loss is useful to sample better tokens and as a result enhances the quality of generated outputs.
\end{itemize}

%% file: src/ch02-relatedwork.tex

\section{Related Work}

\subsection{Generalized Speech Enhancement Models}

In the early days, speech enhancement tasks were often completed by a single model, such as speech denoising~\cite{richter2023sgmse,reddy2020interspeech}, speech dereverberation~\cite{kinoshita2013reverb}, super-resolution~\cite{liu2024audiosr}, target speaker extraction~\cite{wang2024wesep,tang2024tselm}, etc. These tasks are often independent of each other and require different models to be trained separately. In recent years, researchers have proposed some joint framework that integrate multiple enhancement tasks into a single model to improve the model's generalization ability. Liu et al.~\cite{liu2022voicefixer} proposed a new concept called General Speech Restoration (GSR), which aims to solve a wide range of speech enhancement tasks, including denoising, dereverberation, declipping and super-resolution. They introduced the first GSR framework called VoiceFixer, which consists of an ResUNet-based analysis stage and a neural vocoder-based synthesis stage. Li et al.~\cite{li2024masksr} uses a masked generative model whose target is the same as VoiceFixer. Zhang et al.~\cite{zhang2023toward} proposed "Universal SE" model aims at both denoising and dereverberation, and accepts various forms of input(eg. multi-channel, multi-sampling rate). Urgent challenge~\cite{zhang24h_interspeech} focuses on Universality, Robustness, and Generalizability for Speech Enhancement. \todo{However, these models do not include target speaker extraction or consider the singing voice domain.} NeMo~\cite{ku2024generative} and SpeechFlow~\cite{liugenerative} are both flow-matching-based~\cite{lipman2022flow} models that pretrained on a large-scale dataset and can be fine-tuned for downstream enhancement tasks like speech denoising and target speaker extraction. However, these models require fine-tuning on all these downstream tasks. UniAudio~\cite{Yang2024UniAudio} proposed a unified audio foundation model which includes speech denoising, target speaker extraction and can be finetuned to dereverberation task. But this model is not specifically designed for enhancement tasks, and it requires fine-tuning for speech dereverberation and cannot perform super-resolution. 

\subsection{Language Model for Speech Enhancement Tasks}

With the rapid development and widespread adoption of language models, researchers have begun exploring their potential to improve speech enhancement systems. For instance, the SELM model~\cite{wang2024selm} employs a WavLM-based k-means tokenization strategy and utilizes a Transformer-based architecture to realize one-step forward prediction for speech enhancement. Based on SELM, the TSELM model~\cite{tang2024tselm} incorporates a speaker embedding branch to facilitate target speaker extraction. Similarly, Gense~\cite{anonymous2025gense} adopts a two-stage autoregressive paradigm that first maps noisy audio to a semantic space before reconstructing the acoustic signal, then it generates acoustic feature of speech through a semantic-to-acoustic process. In another line of research, MaskSR~\cite{li2024masksr} and Genhancer~\cite{yang2024genhancer} incorporate acoustic tokens from Descript Audio Codec (DAC)~\cite{kumar2024high} and language model to achieve generalized speech restoration capabilities. An extension, MaskSR2~\cite{liu2024joint}, leverages semantic distillation to further improve the intelligibility of the generated speech outputs. The SpeechX model~\cite{wang2024speechx} uses tokens from Encodec~\cite{defossezhigh} and performs multiple speech-related tasks, including text-to-speech, noise suppression, target speaker extraction, and speech editing. However, these existing approaches, except for SpeechX, have not fully harnessed the in-context-learning capabilities of modern language models, and even SpeechX remains limited in addressing certain enhancement tasks, such as dereverberation or super-resolution. 

%% file: src/ch03-method.tex

\section{Background}

\begin{figure*}[htbp]
    \begin{center}
    \includegraphics[width=0.95\linewidth]{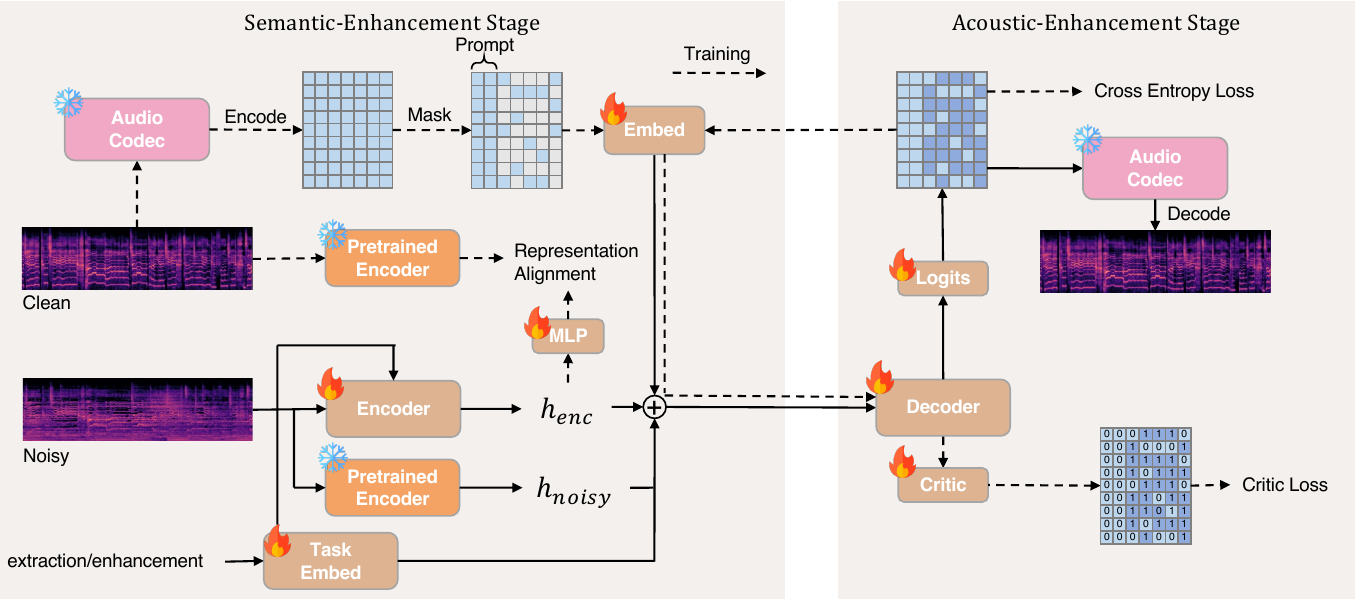}
    \end{center}
    \vspace{-5pt}
    \caption{Model Architecture for \tool. It operates in two stages: (1) the semantic enhancement stage, where the encoder extracts and align semantic features from distorted and clean audio, (2) the acoustic enhancement stage, where the decoder predicts masked tokens using semantic features and acoustic tokens. Part of the clean audio can be retained for prompt-guidance (See Figure~\ref{fig:prompt} for details), and a critic head enables self-critic training and sampling.}
    \label{fig:model}
\end{figure*}

\subsection{Enhancement Problem Formulation}\label{sec:enhancement_problem}
All enhancement problems can be modeled as distortions applied to the original audio. We denote the clean speech/singing audio as $x \in \mathbb{R}^L$, where $L$ is the audio length in sample points. We model the distortion process as a function $d(\cdot)$. The distorted audio $y \in \mathbb{R}^L$ can be written as: $y = d(x)$. We consider a wide range of distortions:

\begin{itemize}[itemsep=1pt,topsep=0pt,parsep=0pt]
    \item Noise: $d_{\text{noise}}(x) = x + n$, where $n$ is the noise signal.
    \item Reverb: $d_{\text{reverb}}(x) = x \ast r$, where $r$ is the room impulse response.
    \item Clipping: $d_{\text{clip}}(x) = max(min(x, \gamma), -\gamma)$, where $\gamma \in [0, 1]$ is the clipping threshold.
    \item Bandwidth Limitation: $d_{\text{bw}}(x) = Resample(x, freq)$, where $freq$ is the target lower sampling frequency.
    \item Other Voice: $d_{\text{voice}}(x) = x + v$, where $v$ is the other voice signal.
\end{itemize}

Previous works~\cite{liu2022voicefixer,zhang24h_interspeech} model distortions in a sequential order. Similarly, we model the overall distortion $D(\cdot)$ as a \textbf{composite function}:
\begin{align*}
    D(x) = d_1(d_2(\ldots d_Q(x)\ldots)), \quad d_q \in \mathbb{D}, \quad q = 1, 2, \ldots, Q.
\end{align*}
Where $\mathbb{D}$ is the set of distortion functions, and $Q$ is the number of distortion types. The goal of our model is to recover the clean audio $x$ from the distorted audio $y$ by learning the inverse function $D^{-1}(\cdot)$. Each distortion function $d_q$ is applied with a probability $p_q$, which is adjusted as a hyperparameter.

\begin{figure}[htbp]
    \begin{center}
        \includegraphics[width=\linewidth]{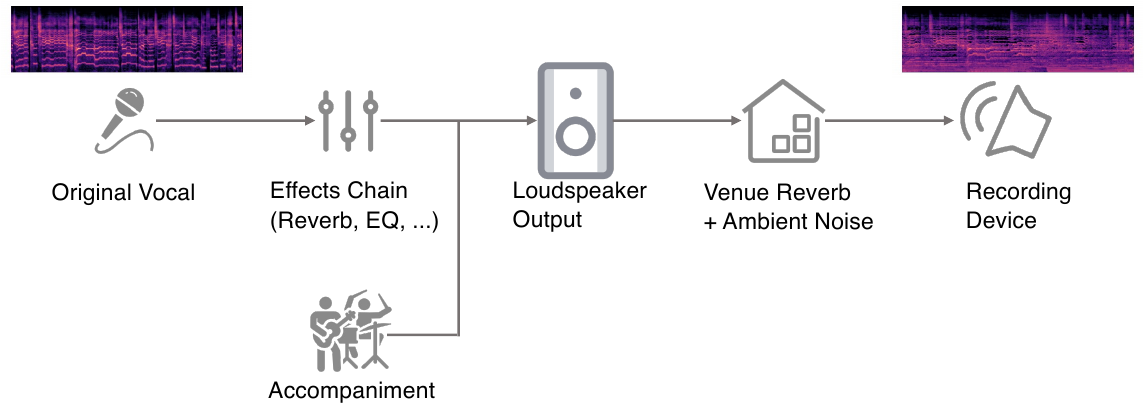}
    \end{center}
    \vspace{-5pt}
    \caption{\majorrevision{Simulation of real-world live vocal recordings, including vocal effects, venue acoustics, and ambient noise.}}
    \label{fig:degradation-formulation}
\end{figure}

When simulating data, previous works typically follow the distortion order of $ d_{\text{noise}}(d_{\text{bw}}(d_{\text{clip}}(d_{\text{reverb}}(\cdot)))) $, placing additive noise at the final stage. However, in real-world scenarios, noise is often intertwined with reverberation and other artifacts. \majorrevision{An overview of the real-world recording path simulated in our framework is illustrated in Figure~\ref{fig:degradation-formulation}.} To better approximate practical conditions, we propose a refined data simulation strategy that applies noise before reverberation, and places bandwidth and clipping effects at the end. Additionally, we insert a separate vocal effects chain (e.g., EQ, reverb) prior to noise to simulate live performance processing. The final degradation chain is formulated as:
\[
    D(x) = d_{\text{bw}}(d_{\text{clip}}(d_{\text{reverb}}(d_{\text{noise}}(d_{\text{vocal\_effect}}(x))))),
\]
where $ d_{\text{vocal\_effect}} = d_{\text{reverb}}(d_{\text{eq}}(x)) $ represents the vocal effect chain. Here, $d_{\text{eq}}$ denotes the equalization effect, which applies 1-3 bell gains in the range of [-5, 5] dB across 10-12,000 Hz within a 1-second window. We define two general tasks for our model: \textbf{Enhancement} and \textbf{Extraction}. The enhancement task aims to recover the clean audio from the distorted audio, while the extraction task aims to extract the target speaker's voice from the mixture. The difference between these two tasks is that the extraction task would insert $d_{\text{voice}}$ into the head of the distortion chain, \majorrevision{namely $D_{extraction}(x) = d_{\text{bw}}(d_{\text{clip}}(d_{\text{reverb}}(d_{\text{voice}}(d_{\text{vocal\_effect}}(x))))).$}

\subsection{Masked Generative Modeling}\label{sec:diffusion}

Masked Generative Modeling can also be formulated as a discrete diffusion process~\cite{lezama2022improved}. Given a discrete token sequence $\mathbf{X}$ of some data. The forward diffusion process at time $t$ is formulated as $\mathbf{X}_t = \mathbf{X} \odot \mathbf{M}_t$, by masking a subset of tokens in $\mathbf{X}$ with the corresponding binary mask $\mathbf{M}_t = [m_{t,i}]_{i=1}^N$. Specifically, we replace $x_i$ with a special [MASK] token if $m_{t,i} = 1$, otherwise keep $x_i$ unmasked if $m_{t,i} = 0$. Each $m_{t,i}$ is i.i.d. according to a Bernoulli distribution with parameter $\gamma(t)$, where $\gamma(t) \in (0, 1]$ represents a mask schedule function (in this paper we use $\gamma(t) = sin(\frac{\pi t}{2T}), t \in (0, T]$). We denote $\mathbf{X}_0 = \mathbf{X}$ as the original token sequence and $\mathbf{X}_T$ as fully masked token sequence. The model is trained to predict the masked tokens based on the unmasked tokens and condition $\mathbf{C}$, modeled as $p_\theta(\mathbf{X}_0 \mid \mathbf{X}_t, \mathbf{C})$. The parameters $\theta$ are optimized to minimize the negative log-likelihood of the masked tokens:
\[
\mathcal{L}_\text{mask} = \mathbb{E}_{\mathbf{X} \in \mathcal{D}, t \in [0, T]}\left[ - \sum_{i=1}^N m_{t,i} \cdot \log\left(p_\theta(x_i\mid\mathbf{X}_t, \mathbf{C})\right) \right].
\]

During inference, the reverse diffusion process iteratively predicts the tokens starting from fully masked token sequence $\mathbf{X}_T$. Suppose the total number of inference steps is $S$, at each timestep $t = \{S, S-1, \dots, 1\}$, we first sample $\hat{\mathbf{X}}_0$ from  the model's logits $p_\theta(\mathbf{X}_0\mid\mathbf{X}_{\frac{t}{S}T},\mathbf{C})$. Then, we choose $ \lfloor N \cdot \gamma(\frac{t-1}{S}T) \rfloor $ tokens to remask based on the confidence score. In common practice~\cite{chang2022maskgit,li2024masksr,wang2024maskgct}, the model's logits $p_\theta(\hat{x}_i\mid\mathbf{X}_{\frac{t}{S}T},\mathbf{C})$ is directly taken as confidence score if $\mathbf{X}_{\frac{t}{S}T, i} = \text{[MASK]}$, otherwise 1, and tokens with the lowest confidence scores are remasked.

\section{Proposed Unified Model with Prompt Guidance and Self-Critic}

\subsection{Model Architecture}\label{sec:model_architecture}
The proposed model architecture is shown in Figure~\ref{fig:model}. Similar to MaskSR~\cite{li2024masksr,liu2024joint}, our model is also a Transformer-based Masked Generative Model. The model is divided into two stages: the semantic enhancement stage and the acoustic enhancement stage. In the semantic enhancement stage, the encoder is responsible for extracting the semantic features from the input distorted audio using representations aligned with pre-trained features. In the acoustic enhancement stage, the decoder predicts masked tokens based on the semantic features and existing acoustic tokens. A critic head is introduced to perform self-critic training and sampling.

We use pretrained Descript Audio Codec (DAC)~\cite{kumar2024high} as audio tokenizer to model clean voice. Through 9 residual vector quantizer layers, we obtain a sequence of tokens of shape $\mathbf{X}_0 \in \mathbb{R}^{9 \times T} $ of clean audio $x$. At timestep $t$, we mask a subset of tokens in $\mathbf{X}_0$ to obtain $\mathbf{X}_t$. To deal with residual layers, we employ 9 embedding layers for each residual layer and sum over the residual layers to obtain the token latent $h_{\text{token}} \in \mathbb{R}^{T \times \text{dim}}$.

During the semantic enhancement stage, the input of the encoder is the power-law compressed magnitude STFT spectrogram~\cite{R14} of the distorted audio $D(x)$ and task embedding $h_{\text{task}}$. The encoder consists of $L_{enc}$ self-attention transformer layers to obtain the latent representation $h_{\text{enc}} \in \mathbb{R}^{T \times dim}$. Inspired by representation alignment~(REPA) in the field of image generation~\cite{yu2024representation} and semantic distillation in MaskSR2~\cite{liu2024joint}, we introduce a representation alignment module to align the encoder output. The semantic feature of the clean audio is extracted using self-supervised learning (SSL) models, such as w2v-BERT~\cite{chung2021w2v}. Let $f$ be a SSL encoder, $z = f(x) \in \mathbb{R}^{T \times dim}$ be an encoder output~(linear interpolation is used to match time dimension $T$ and feature dimension $dim$), a multilayer perceptron $\text{MLP}_\phi$ is used to project the encoder output with the clean audio's semantic feature:

\[
\mathcal{L}_{\text{REPA}} = \mathbb{E}_{x} \left[-\sum_{t=1}^T \text{sim}\left(z^{[t]}, \text{MLP}_\phi(h_{\text{enc}}^{[t]})\right) \right],
\]

where the similarity function is mean-square error (MSE). \majorrevision{Furthermore, motivated by~\cite{hung2022boosting,huang2022investigating}, we also integrate self-supervised feature to the output for semantic enhancement. Prior work has shown that SSL models possess inherent noise robustness and can significantly improve speech enhancement performance when used as input features.} Based on this insight, the final output of our semantic enhancement stage is formulated as the sum of the encoder output and the noisy semantic feature: $h_{\text{semantic}} = h_{\text{enc}} + f(D(x))$.

During the acoustic enhancement stage, discrete diffusion process is employed. The condition consists of semantic features and task embeddings $\mathbf{C} = \{h_{\text{semantic}}, h_{\text{task}}\}$. The decoder consists of $L_{dec}$ self-attention transformer layers. Finally, 9 linear layers project the output of the last decoder layer back to logits of each residual layers. $\mathcal{L}_{\text{mask}}$ is computed between the predicted masked token and the target token.

\majorrevision{Though \tool adopts MaskSR's architecture, it introduces several key innovations, including degradation path optimization, prompt-guidance and self-critic sampling. These concepts are generally applicable and beneficial beyond the MaskSR framework. While MaskSR focuses solely on speech restoration, \tool extends its applicability to multiple tasks such as target speech extraction and singing voice enhancement. Our prompt-guidance mechanism is not limited to enabling the TSE task. Inspired by the timbre cloning capability in zero-shot TTS, we aim to leverage prompt guidance to enhance timbre similarity and overall quality across a broader range of tasks.}

\subsection{Prompt Guidance}
\begin{figure}[htbp]
    \begin{center}
        \includegraphics[width=\linewidth]{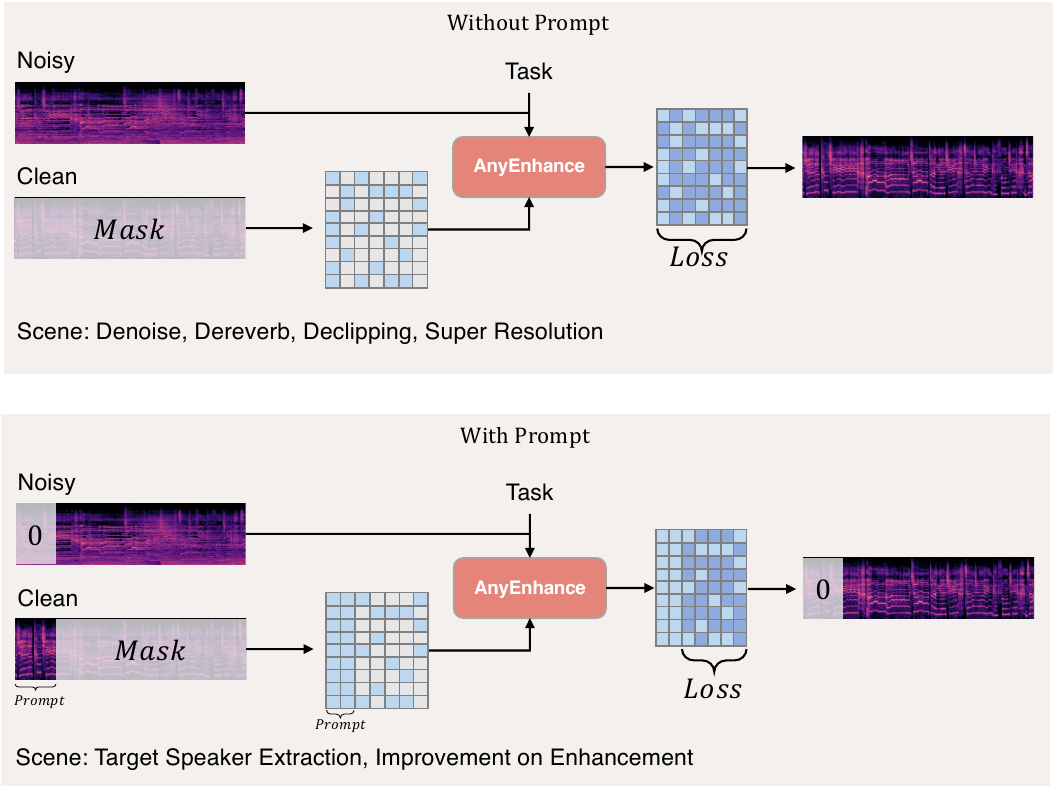}
    \end{center}
    \vspace{-5pt}
    \caption{Comparison of model training with and without prompts. The inclusion of prompts enables the model to naturally perform TSE also achieving improved performance across other enhancement tasks.}
    \label{fig:prompt}
\end{figure}

We introduce a prompt-guidance scheme (see Figure~\ref{fig:prompt}). \majorrevision{The design of the prompt-guidance mechanism is inspired by the success of zero-shot TTS systems~\cite{ju2024naturalspeech,wang2024maskgct}, where models can mimic the timbre and audio characteristics of a reference speaker through prompt conditioning. Our goal is to extend this capability to speech enhancement tasks, enabling the model to leverage clean prompt audio to learn target speaker characteristics and audio quality in an end-to-end manner without alternating the model architecture.} The model can be trained using both configurations: without prompt and with prompt. During training, we have a probability $p_{\text{prompt}}$ to keep the first 3 seconds of clean audio unmasked as the prompt. The same place of the distorted audio is set to silence. The prompt section is not considered for loss calculation. During inference, the model can infer both with a prompt audio and without prompt audio:

\begin{itemize}[itemsep=1pt,topsep=0pt,parsep=0pt]
    \item Without a prompt, the clean token is set to a fully masked state, and the model starts restoration from scratch.
    \item With a prompt, the prompt token is appended to the beginning of the clean code, while the rest of the sequence remains fully masked, guiding the model to generate the desired output with target speaker's voice as reference.
\end{itemize}

The prompt-guidance scheme ensures the model to seamlessly accept a reference speaker's voice for target speaker extraction and also improves performance on other tasks when a reference speaker's prompt audio is proveded.

\subsection{Self-Critic Sampling}
\label{sec:self-critic}

The sampling strategy described in Section~\ref{sec:diffusion} is based on a greedy approach, which indicates that the sampling process is non-regrettable, thus cannot recover already filled tokens. Moreover, the confidence score of each token is determined independently (by simply taking the model's logits as the confidence score). To address these issues, we introduce Self-Critic sampling strategy.

The core idea of critic-sampling is to use a model as a critic or discriminator to evaluate whether a token is real or generated by the model. After we first sample $\hat{\mathbf{X}}_0$ from $p_\theta(\mathbf{X}_0\mid\mathbf{X}_t,\mathbf{C})$ and the model's logits, the unmasked tokens in $ \mathbf{X}_t $ are copied into the output to form $\tilde{\mathbf{X}}_0 = \hat{\mathbf{X}}_0 \odot \mathbf{M}_t + \mathbf{X}_0 \odot (1-\mathbf{M}_t)$. The critic model (parameterized by $\phi$) is trained to estimate $\mathbf{M}_t$ from $\tilde{\mathbf{X}}_0$ using binary cross-entropy loss:
\[
    \mathcal{L}_\text{critic} = \mathbb{E}_{\substack{\mathbf{X} \in \mathcal{D}, t \in [0, T], \\ p_\theta(\tilde{\mathbf{X}}_0\mid\mathbf{X}_t,\mathbf{C})}}\left[ - \sum_{i=1}^N m_{t,i} \cdot \log\left(p_{\phi }(\hat{m}_{t,i}\mid\tilde{\mathbf{X}}_0, \mathbf{C})\right) \right].
\]
    
For Token-Critic~\cite{lezama2022improved}, a separate critic model $\phi$ is trained. Inspired by SCRIPT~\cite{nijkamp2021script}, which was originally proposed for pretraining masked language models such as BERT~\cite{devlin2019bert}, we use the generative model itself as the critic model. We introduce an additional linear layer, referred to as the \textit{\textbf{critic head}}, to replace the model's logits projection layer. During training, we feed $\tilde{\mathbf{X}}_0$ back into the model itself with the critic head to produce the critic score $\hat{\mathbf{M}}_{t}$ and minimize the binary cross-entropy loss between $\mathbf{M}_t$. Self-Critic eliminates the need for training a separate critic model and still improves the masked generative model's performance by providing a more accurate confidence score. The final loss is:
\[
\mathcal{L} = \mathcal{L}_\text{mask} + \mathcal{L}_\text{REPA} + \mathcal{L}_\text{critic}.
\]

%% file: src/ch04-experiments.tex

\section{Experiments}

\subsection{Datasets}

\begin{table}[t]
    \centering
    \caption{Statistics of the speech, singing voice, noise and room impulse response (RIR) datasets used for \tool.}
    \label{tab:dataset_statistics}
    \begin{tabular}{p{1cm}p{4.2cm}l}
        \toprule
        \textbf{Category} & \textbf{Datasets} & \textbf{Size}\\
        \midrule
        24kHz Speech & Emilia~\cite{he2024emilia} (Filtered Subset) & 20.987.49 hours \\
        \midrule
        Fullband Speech & VCTK~\cite{yamagishi2019vctk}, HiFi-TTS~\cite{bakhturina2021hi}, HQ-TTS~\cite{liu2022voicefixer}, AI-Shell3~\cite{AISHELL3}, Bible-TTS~\cite{meyer2022bibletts} & 956.74 hours \\
        \midrule
        Fullband Singing & OpenSinger~\cite{OpenSinger}, PopCS~\cite{diffsinger}, PopBuTFy~\cite{popbutfy}, Opencpop~\cite{Opencpop}, M4Singer~\cite{M4Singer}, SingStyle111~\cite{SingStyle}, ACESinger~\cite{ACESinger} & 572.01 hours \\
        \midrule
        Noise & MUSAN~\cite{snyder2015musan}, Urgent Challenge~\cite{zhang24h_interspeech}, FSD50K~\cite{fonseca2021fsd50k}, DESED~\cite{turpault2019sound}, TUT Urban Acoustic Scenes~\cite{mesaros2018multi}, Live-performance Noise & 1233.89 hours \\
        \midrule
        RIR & SLR26, SLR28, Self-collected & 62,668 pieces \\
        \bottomrule
    \end{tabular}
\end{table}

\textbf{Training Datasets}  
The training datasets used in our experiments include speech, singing voice, noise, and room impulse responses (RIRs), with detailed statistics shown in Table~\ref{tab:dataset_statistics}. For speech data, we used 24kHz speech from Emilia~\cite{he2024emilia} Subset (filtered with DNSMOS OVRL $>$ 3.4, totaling $\sim$21k hours) and fullband speech from datasets such as VCTK~\cite{yamagishi2019vctk}, HiFi-TTS~\cite{bakhturina2021hi} ($\sim$1k hours), and AI-Shell3~\cite{AISHELL3}. Singing voice data include OpenSinger~\cite{OpenSinger}, PopCS~\cite{diffsinger}, and others ($\sim$600 hours). Noise data are sourced from the musan noise subset~\cite{snyder2015musan}, Urgent Challenge noise data~\cite{zhang24h_interspeech}, FSD50K~\cite{fonseca2021fsd50k}, DESED~\cite{turpault2019sound}, and the TUT Urban Acoustic Scenes 2018 dataset~\cite{mesaros2018multi}, along with a proprietary dataset of live-performance noise ($\sim$1.2k hours). RIRs comprise 62,668 pieces from SLR26, SLR28, and self-collected RIRs from various environments. We dynamically distort the voice with noise and RIR data to get the training data pair for the Enhancement and Extraction tasks described in Section~\ref{sec:enhancement_problem}. All the audio pieces are resampled to 44.1kHz. Table~\ref{tab:distortion_prob} shows the probability and hyperparameters of the distortion used for training data.

\begin{table}[htbp]
    \centering
    \caption{Probability and hyperparameters of the distortion used for training data (See Section~\ref{sec:enhancement_problem} for detailed definition).}
    \label{tab:distortion_prob}
    \begin{tabular}{lcp{4.8cm}}
        \toprule
        \textbf{Distortion} & \textbf{Probability} & \textbf{Hyperparameters} \\
        \midrule
        Noise & 0.9 & Signal-to-noise ratio (SNR) $\in$ [-5, 20] dB. \\
        \midrule
        Reverb & 0.5 & - \\
        \midrule
        Clipping & 0.25 & Clipping threshold $\in$ [0.06, 0.9]. \\
        \midrule
        \makecell[l]{Bandwidth\\Limitation} & 0.5 & Bandwidth $\in$ \{2, 4, 8, 16, 24, 32\}kHz. \\
        \midrule
        Other Voice & 0.5 & Signal-to-noise ratio (SNR) $\in$ [0, 10] dB. \\
        \midrule
        Vocal Effect & 0.5 & $d_{\text{vocal\_effect}} = d_{\text{reverb}} \circ d_{\text{eq}}$. \\
        \bottomrule
    \end{tabular}
\end{table}

\textbf{Evaluation Datasets} We evaluate the performance of \tool across multiple speech and singing evaluation datasets, grouped by types of distortions and corresponding tasks: General Speech Restoration (GSR) Group, Speech Enhancement (SE) Group, Super Resolution (SR) Group, and Target Speaker Extraction (TSE) Group. Detailed information is as follows:

\begin{itemize}[itemsep=1pt,topsep=0pt,parsep=0pt]
    \item \textbf{GSR Group}: We use the official Voicefixer GSR dataset~\cite{liu2022voicefixer} and two simulated GSR test sets based on Librivox~\cite{kearns2014librivox} (speech) and CCMusic acapella~\cite{ccmusic} (singing) to evaluate the ability of the model to handle simultaneous distortions, including noise, reverberation, clipping, and bandwidth limitation.
    \item \textbf{SE Group}: Focuses on the traditional Speech Enhancement (SE) tasks. We use the official 2020 DNS Challenge test sets~\cite{reddy2020interspeech}, specifically two simulated test sets with and without reverb, evaluating the ability of the model to handle distortions such as noise and reverberation.
    \item \textbf{SR Group}: Focuses on the Super Resolution (SR) Task, only addressing bandwidth limitation. We use the official Voicefixer SR subset (speech) and the simulated CCMusic acapella SR test set (singing). Input bandwidths range from 2kHz to 8kHz.
    \item \textbf{TSE Group}: Focuses on the Target Speaker Extraction Task. We use the official Librimix~\cite{cosentino2020librimix} test set which only includes other voice as distortion, and a simulated VCTK~\cite{yamagishi2019vctk} TSE test set which incorporates additional distortions such as noise, reverberation, clipping, and bandwidth limitation beyond other voices.
\end{itemize}

\begin{table*}[!ht]
    \centering
    \scriptsize
    \caption{Performance summary of \tool across various enhancement datasets. *Indicates that the evaluation dataset is sampled at 16 kHz.}
    \label{tab:result-main}
    \renewcommand\arraystretch{1.4}
    \begin{tabular}{
        |l
        |p{2cm}
        |l
        |l
        |l
        |cccccc|}
        \toprule
        \textbf{Group} & \textbf{Distortion(s)} & \textbf{Domain} & \textbf{Dataset} & \textbf{Model} & \textbf{SIG$\uparrow$} & \textbf{BAK$\uparrow$} & \textbf{OVRL$\uparrow$} & \textbf{NISQA$\uparrow$} & \makecell{\textbf{Speech-}\\\textbf{BERTScore$\uparrow$}} & \textbf{Similarity$\uparrow$} \\
        \cline{1-11}
        \multirow{20}{*}{GSR} &
        \multirow{20}{*}{\makecell[l]{Noise,\\Reverb,\\Clipping,\\Bandwidth\\Limitation}} &
        \multirow{14}{*}{Speech} &
        \multirow{6}{*}{Voicefixer GSR}
        & TFGridNet & 3.253 & 3.906 & 2.945 & 3.643 & 0.782 & 0.613 \\
        & & & & NSNet2 & 3.011 & 3.969 & 2.758 & 3.433 & 0.728 & 0.615 \\
        & & & & Voicefixer (checkpoint) & 3.299 & 3.971 & 3.003 & 4.16 & 0.797 & 0.882 \\
        & & & & Voicefixer (retrained) & 3.3 & 3.984 & 2.996 & 4.054 & 0.818 & 0.884 \\
        & & & & MaskSR & \textbf{3.408} & 4.041 & 3.122 & \textbf{4.335} & \textbf{0.832} & 0.916 \\
        \cline{5-11}
        & & & & \tool & 3.406 & \textbf{4.073} & \textbf{3.136} & 4.308 & 0.829 & \textbf{0.924} \\
        \cline{4-11}
        & & & \multirow{6}{*}{Librivox GSR}
        & TF-GridNet & 3.274 & 3.872 & 2.951 & 3.138 & 0.77 & 0.931 \\
        & & & & NSNet2 & 2.895 & 3.866 & 2.589 & 2.735 & 0.7 & 0.892 \\
        & & & & Voicefixer (checkpoint) & 3.365 & 3.971 & 3.085 & 3.77 & 0.706 & 0.864 \\
        & & & & Voicefixer (retrained) & 3.35 & 4.024 & 3.069 & 3.63 & 0.758 & 0.897 \\
        & & & & MaskSR & 3.499 & 4.133 & 3.258 & 4.155 & 0.779 & 0.94 \\
        \cline{5-11}
        & & & & \tool & \textbf{3.546} & \textbf{4.142} & \textbf{3.308} & \textbf{4.346} & \textbf{0.822} & \textbf{0.955} \\
        \cline{3-11}
        & & \multirow{6}{*}{Singing} & 
        \multirow{6}{*}{CCMusic GSR}
        & TF-GridNet & 2.764 & 3.37 & 2.362 & 2.396 & 0.57 & 0.81 \\
        & & & & NSNet2 & 2.608 & 3.645 & 2.226 & 2.439 & 0.574 & 0.798 \\
        & & & & Voicefixer (checkpoint) & 2.75 & 3.094 & 2.354 & 2.917 & 0.636 & 0.823 \\
        & & & & Voicefixer (retrained) & 2.948 & 3.349 & 2.551 & 3.224 & 0.738 & 0.881 \\
        & & & & MaskSR & 3.153 & 3.483 & 2.715 & 3.157 & 0.77 & 0.889 \\
        \cline{5-11}
        & & & & \tool & \textbf{3.243} & \textbf{3.547} & \textbf{2.797} & \textbf{3.345} & \textbf{0.811} & \textbf{0.915} \\
        \cline{1-11}
        \multirow{23}{*}{SE} &
        \multirow{11}{*}{\makecell[l]{Noise}} &
        \multirow{23}{*}{Speech} &
        \multirow{11}{*}{DNS No Reverb$^{*}$}
        & DEMUCS & 3.533 & 4.157 & 3.31 & 3.742 & 0.877 & 0.984 \\
        & & & & FRCRN & 3.574 & 4.154 & 3.332 & 4.495 & \textbf{0.914} & \textbf{0.993} \\
        & & & & SGMSE & 3.501 & 3.710 & 3.137 & - & - & - \\
        & & & & StoRM & 3.514 & 3.941 & 3.205 & - & - & - \\
        & & & & SELM & 3.508 & 4.096 & 3.258 & - & - & - \\
        & & & & TFGridNet & 3.539 & 4.047 & 3.268 & 4.347 & 0.902 & 0.675 \\
        & & & & NSNet2 & 3.391 & 4.071 & 3.132 & 3.943 & 0.847 & 0.675 \\
        & & & & Voicefixer (checkpoint) & 3.504 & 4.109 & 3.253 & 4.274 & 0.819 & 0.956 \\
        & & & & Voicefixer (retrained) & 3.452 & 4.035 & 3.157 & 4.105 & 0.827 & 0.959 \\
        & & & & MaskSR & 3.616 & \textbf{4.183} & 3.393 & 4.754 & 0.875 & 0.983 \\
        \cline{5-11}
        & & & & \tool & \textbf{3.64} & 4.179 & \textbf{3.418} & \textbf{4.821} & 0.907 & 0.988 \\
        \cline{2-2}
        \cline{4-11}
        & \multirow{11}{*}{\makecell[l]{Noise,\\Reverb}} & & 
        \multirow{11}{*}{DNS With Reverb$^{*}$}
        & DEMUCS & 2.937 & 3.844 & 2.615 & 2.188 & 0.725 & 0.93 \\
        & & & & FRCRN & 2.933 & 2.923 & 2.279 & 2.27 & 0.783 & \textbf{0.966} \\
        & & & & SGMSE & 2.730 & 2.741 & 2.430 & - & - & - \\
        & & & & StoRM & 2.947 & 3.141 & 2.516 & - & - & - \\
        & & & & SELM & 3.160 & 3.577 & 2.695 & - & - & - \\
        & & & & TFGridNet & 3.11 & 3.225 & 2.51 & 2.614 & \textbf{0.84} & 0.686 \\
        & & & & NSNet2 & 2.756 & 3.719 & 2.421 & 2.043 & 0.763 & 0.691 \\
        & & & & Voicefixer (checkpoint) & 3.43 & 4.016 & 3.132 & \textbf{3.822} & 0.711 & 0.91 \\
        & & & & Voicefixer (retrained) & 3.074 & 3.721 & 2.667 & 2.906 & 0.724 & 0.918 \\
        & & & & MaskSR & 3.396 & \textbf{4.043} & 3.085 & 3.353 & 0.701 & 0.946 \\
        \cline{5-11}
        & & & & \tool & \textbf{3.5} & 4.04 & \textbf{3.204} & 3.722 & 0.738 & 0.951 \\
        \cline{1-11}
        \multirow{11}{*}{SR} &
        \multirow{11}{*}{\makecell[l]{Bandwidth\\Limitation}} &
        \multirow{5}{*}{Speech} &
        \multirow{5}{*}{Voicefixer SR}
        & Voicefixer (checkpoint) & 3.405 & 4.029 & 3.11 & 4.131 & 0.873 & 0.882 \\
        & & & & Voicefixer (retrained) & 3.041 & 3.903 & 2.745 & 3.556 & 0.837 & 0.854 \\
        & & & & AudioSR & \textbf{3.492} & 4.002 & 3.18 & 4.255 & 0.913 & 0.911 \\
        & & & & MaskSR & 3.464 & 4.028 & 3.154 & \textbf{4.352} & 0.925 & 0.939 \\
        \cline{5-11}
        & & & & \tool & 3.449 & \textbf{4.063} & \textbf{3.156} & 4.201 & \textbf{0.941} & \textbf{0.943} \\
        \cline{3-11}
        & & \multirow{5}{*}{Singing} &
        \multirow{5}{*}{CCMusic SR}
        & Voicefixer (checkpoint) & 3.179 & 3.534 & 2.743 & \textbf{3.356} & 0.463 & 0.65 \\
        & & & & Voicefixer (retrained) & 3.108 & 3.504 & 2.692 & 3.218 & 0.768 & 0.864 \\
        & & & & AudioSR & 3.192 & 3.531 & 2.75 & 2.836 & 0.468 & 0.63 \\
        & & & & MaskSR & 3.308 & 3.588 & 2.857 & 3.173 & 0.813 & 0.892 \\
        \cline{5-11}
        & & & & \tool & \textbf{3.339} & \textbf{3.628} & \textbf{2.899} & 3.225 & \textbf{0.854} & \textbf{0.919} \\
        \cline{1-11}
        \multirow{7}{*}{TSE} &
        \multirow{3}{*}{\makecell[l]{Other Voice}} &
        \multirow{7}{*}{Speech} &
        \multirow{3}{*}{Librimix TSE$^{*}$}
        & WeSep & 3.563 & 3.931 & 3.228 & 4.041 & \textbf{0.922} & \textbf{0.991} \\
        & & & & TSELM & 3.55 & \textbf{4.084} & 3.288 & 4.029 & 0.808 & 0.908 \\
        \cline{5-11}
        & & & & \tool & \textbf{3.638} & 4.066 & \textbf{3.353} & \textbf{4.277} & 0.735 & 0.914 \\
        \cline{2-2}
        \cline{4-11}
        & \multirow{3}{*}{\makecell[l]{Other Voice, Noise,\\Reverb, Clipping,\\Bandwidth Limita-\\tion}} & &
        \multirow{3}{*}{VCTK TSE}
        & WeSep & 2.483 & 2.191 & 1.933 & 1.959 & 0.568 & 0.856 \\
        & & & & TSELM & 3.345 & 3.875 & 3.004 & 3.388 & 0.58 & 0.81 \\
        \cline{5-11}
        & & & & \tool & \textbf{3.545} & \textbf{4.102} & \textbf{3.275} & \textbf{4.57} & \textbf{0.727} & \textbf{0.925} \\
        \bottomrule
    \end{tabular}
\end{table*}

\subsection{Baseline Systems}

We compare the performance of \tool with several state-of-the-art models on the GSR, SE, SR, and TSE Groups. The details of the baseline models are as follows:

\textbf{GSR Group Baselines}: We evaluate the performance of \tool on the GSR Group by comparing it with TF-GridNet~\cite{wang2023tf}, a time-frequency domain multi-path deep neural network provided as the 2024 Urgent Challenge~\cite{zhang24h_interspeech} baseline; NSNet2~\cite{braun2020data}, a regression model provided as the 2021 DNS Challenge\cite{reddy2021icassp} baseline; Voicefixer~\cite{liu2022voicefixer}, a ResUNet-based two-stage speech restoration model and MaskSR~\cite{li2024masksr}, a masked generative model for speech restoration tasks. We use the official checkpoints provided by the authors for TF-GridNet, NSNet2. For Voicefixer, we use the official checkpoint provided by the authors and also retrain the model on our datasets to ensure a fair comparison. For MaskSR we reproduced the model ourselves and set the parameters to be the same as \tool. It is worth noting that these GSR baselines can also be evaluated on the SE task, and both Voicefixer and MaskSR are also evaluated on the SR task as well.

\textbf{SE Group Baselines}: We evaluate the performance of \tool on the SE Group by comparing it with the aforementioned GSR baselines and several state-of-the-art models designed for speech enhancement, including \todo{DEMUCS~\cite{defossez2019demucs}, FRCRN~\cite{zhao2022frcrn}, SGMSE~\cite{richter2023sgmse}, StoRM~\cite{lemercier2023storm}, and SELM~\cite{wang2024selm}.} We use official checkpoints for DEMUCS and FRCRN, and results for SGMSE, StoRM, and SELM are taken from~\cite{wang2024selm}.

\textbf{SR Group Baselines}: We evaluate the performance of \tool on the SR Group by comparing it with Voicefixer, MaskSR and AudioSR~\cite{liu2024audiosr}, a latent diffusion model designed for super resolution. We use the official checkpoint provided by the authors for AudioSR.

\textbf{TSE Group Baselines}: We evaluate the performance of \tool on the TSE Group by comparing it with the WeSep~\cite{wang2024wesep} TSE checkpoint provided by the toolkit, and TSELM~\cite{tang2024tselm}, a language-model-based model for TSE. We use the official checkpoints provided by the authors for WeSep and TSELM.

\subsection{Implementation Details}

We use the pretrained 44.1kHz DAC~\cite{kumar2024high} codec as our audio tokenizer. For the main experiments, we use a configuration of $ dim=768, L_{enc}=24, L_{dec}=24 $ as configured in Section~\ref{sec:model_architecture} (768 hidden dim, 24 transformer layers for Encoder, 24 transformer layers for Decoder in Figure~\ref{fig:model}), totaling of 363.54M parameters, designed to comprehensively evaluate the model’s enhancement capabilities for both speech and singing voice, utilizing the full dataset and the comprehensive data degradation pipeline. We use rotary position embedding RoPE~\cite{su2024roformer} for the transformer layers as observed in~\cite{zhang2024empirical}. We employ the 17th layer of w2v-bert2\footnote{\url{https://huggingface.co/facebook/w2v-bert-2.0}} for semantic feature alignment. Each audio sample is 8 seconds long, with a 50\% probability of retaining the first 3 seconds as the prompt. Training is conducted on 4 GPUs, with a batch size of 8 per GPU using Adam optimizer for 300k steps, with a learning rate of 1e-4, a 4k-step warmup and linear decay to 0.

For the task and distortion chain ablation studies, we use a smaller model with $ dim=512, L_{enc}=8, L_{dec}=12 $, totaling 77.93M parameters. The model is trained on the 44.1kHz speech dataset and a subset of noise data totaling 400 hours. Each audio sample is 6 seconds long. Training is conducted on 2 GPUs, with a batch size of 16 per GPU. The rest of the settings remain the same as in the main experiment.

\subsection{Evaluation Metrics}

To comprehensively evaluate the performance of \tool, we adopt the following objective and subjective metrics, which are designed to assess various aspects of enhancement quality, including perceptual quality, semantic similarity and speaker similarity:

\textbf{DNSMOS~\cite{reddy2022dnsmos}}: a reference-free perceptual quality estimator that works with 16 kHz audio signals. It outputs 3 scores ranging from 1 to 5: \textbf{SIG}, \textbf{BAK}, and \textbf{OVRL}, which represent the signal quality, background noise, and overall quality, respectively.

\textbf{NISQA~\cite{mittag2021nisqa}}: a reference-free perceptual quality estimator that works with 48 kHz audio signals. It outputs a single score ranging from 1 to 5, which represents the overall quality of the audio signal.

\textbf{SpeechBERTScore~\cite{saeki2024speechbertscore}}: calculates BERTScore on dense self-supervised speech features, which is designed to evaluate the semantic similarity between the enhanced and the reference signals. Here we utilize the pretrained HuBERT-base\footnote{\url{https://huggingface.co/facebook/hubert-base-ls960}} model to extract semantic features.

\textbf{Similarity}: calculates the speaker cosine similarity between the enhanced and the reference signals. We use the pretrained WavLM\footnote{\url{https://huggingface.co/microsoft/wavlm-base-plus-sv}} model to extract speaker embeddings.

\textbf{MOS}: we also conducted listening tests to get the mean opinion score (MOS) of the enhanced audio signals on the Librivox GSR dataset and the CCMusic GSR dataset. We enrolled 10 participants for each dataset, and each participant listened to 70 samples. The MOS is calculated on a 5-point scale ranging from 1 to 5. 

Some common signal-based metrics, such as PESQ~\cite{pesq2}, STOI~\cite{STOI}, and ESTOI~\cite{ESTOI}, are often used to evaluate enhancement models. However, these metrics are unsuitable for generative-model-based enhancement models due to the lack of waveform-level alignment between the enhanced and reference signals~\cite{li2024masksr,wang2024selm,tang2024tselm}. \majorrevision{As a result, we do not include these metrics in our main evaluations and instead focus on perceptual and semantic-level metrics that better reflect the quality of generative outputs. Nevertheless, for a comprehensive evaluation and analysis, we still report these alignment-dependent metrics, as well as the Word Error Rate (WER), in Table~\ref{tab:result-wer-pesq} for the GSR Group.}

\subsection{Experiment Results}

\subsubsection{Evaluation on Various Enhancement Tasks}

We evaluate the performance of \tool across multiple speech and singing evaluation datasets on different tasks. See Table~\ref{tab:result-main}, the performance of \tool is consistently superior or competitive across all tasks and datasets. In GSR Group, for the Voicefixer GSR dataset, \tool achieves the highest OVRL (3.136), NISQA (4.308), and similarity score (0.924), surpassing all baselines. Similarly, on the Librivox GSR dataset, \tool demonstrates the best results in SIG (3.546), OVRL (3.308), and Similarity (0.955), highlighting its robustness. On the CCMusic GSR dataset, \tool outperforms other methods in NISQA (3.345) and achieves competitive OVRL (2.797) and similarity (0.915), further demonstrating its effectiveness under singing voice domain.

In the SE Group, \tool demonstrates competitive performance across both DNS No Reverb and DNS With Reverb datasets. For DNS No Reverb, \tool achieves the highest scores in OVRL (3.418), NISQA (4.821), and Similarity (0.988), establishing its robustness in speech denoising. For DNS With Reverb, \tool achieves the highest scores in SIG (3.500) and OVRL(3.204), and second-highest scores in BAK, NISQA and Similarity. Unlike most baseline models that are limited to 16 kHz outputs, \tool supports full-band restoration, significantly enhancing its applicability to high-quality audio tasks.

For the SR Group, \tool achieves the highest BAK, OVRL, SpeechBERTScore and Similarity scores on both Voicefixer SR and CCMusic SR dataset, and highest SIG score on Voicefixer SR dataset, outperforming all baselines.

Additionally, in the TSE Group, \tool delivers competitive results, achieving the highest scores across SIG, OVRL, NISQA metrics on both Librimix TSE and VCTK TSE datasets. Particularly excelling in challenging noisy scenarios (VCTK TSE) - achieving highest scores across all metrics. The VCTK TSE dataset, with its complex simultaneous distortions, further highlights the limitations of current single-task models, which lack the multitask capability required to handle such diverse challenges effectively.

Overall, \tool exhibits robust generalization and effectiveness across diverse enhancement tasks in both speech and singing domains, outperforming or matching state-of-the-art baselines across various metrics.

\begin{table}[htbp]
    \centering
    \caption{MOS Scores for Different Models on GSR Datasets}
    \label{tab:mos_scores}
    \begin{tabular}{lcc}
        \toprule
        \textbf{Model} & \textbf{Librivox GSR} & \textbf{CCMusic GSR} \\
        \midrule
        Clean & 4.833 & 4.750 \\
        \midrule
        TFGridNet & 2.867 & 2.650 \\
        NSNet2 & 2.483 & 2.417 \\
        Voicefixer (checkpoint) & 2.983 & 2.433 \\
        Voicefixer (retrained) & 2.950 & 3.133 \\
        MaskSR & 4.033 & 3.817 \\
        \midrule
        \tool & \textbf{4.183} & \textbf{3.917} \\
        \bottomrule
    \end{tabular}
\end{table}

Table \ref{tab:mos_scores} presents the MOS scores for different models on the Librivox GSR and CCMusic GSR datasets. \tool achieves the highest MOS scores on both datasets, demonstrating its superior performance in enhancing the listening experience of distorted audio signals based on human evaluations.

\begin{table}[h]
    \centering
    \caption{Comparison With Closed-Source Models On Voicebank-Demand Dataset, results are taken from~\cite{ku2024generative}, PESQ, ESTOI are signal-based metrics and WV-MOS is a reference-free perceptual quality like DNSMOS.}
    \label{tab:close_source_comparison}
    \renewcommand{\arraystretch}{1.2}
    \begin{tabular}{lccc}
    \toprule
    \textbf{Model} & \textbf{WV-MOS$\uparrow$} & \textbf{PESQ$\uparrow$} & \textbf{ESTOI$\uparrow$} \\
    \midrule
    SpeechFlow & - & 3.13 & 0.87 \\
    Nemo & 4.41 & \textbf{3.27} & \textbf{0.88} \\
    \midrule
    \tool & \textbf{4.44} & 2.78 & 0.84 \\
    \bottomrule
    \end{tabular}
\end{table}

Table~\ref{tab:close_source_comparison} provides a comparison between \tool and closed-source models on the Voicebank-Demand dataset. \tool achieves the highest WV-MOS score, while Nemo outperforms \tool in PESQ and ESTOI due to the inherent limitations of generative models in terms of signal-based metrics. These results demonstrate the effectiveness of \tool in enhancing the quality of speech signals compared to closed-source models.

\majorrevision{Additionally, we report the intelligibility metric (Word Error Rate, WER) and signal-based metrics (PESQ, ESTOI) on the Voicefixer GSR, Librivox GSR, and CCMusic GSR datasets in Table~\ref{tab:result-wer-pesq}. We use Whisper-large-v3\footnote{\url{https://huggingface.co/openai/whisper-large-v3}}, a state-of-the-art ASR model, for WER evaluation. However, it sometimes fails to transcribe even the ground-truth audios in the CCMusic GSR test set, resulting in unreliable WER scores. So we do not report WER for this dataset. The results show that \tool consistently achieves competitive or superior performance across all three datasets. On Voicefixer GSR \tool ranks second in WER, PESQ, and ESTOI, close behind TFGridNet. On Librivox GSR, \tool yields the lowest WER (0.199), the second-highest PESQ (2.073), and the best ESTOI (0.784). On the singing voice domain (CCMusic GSR), \tool significantly surpasses all other systems in both PESQ (2.337) and ESTOI (0.585), showing its strong generalization ability in singing-voice-oriented enhancement scenarios.}

\begin{table}[h]
    \centering
    \scriptsize
    \caption{\majorrevision{Intelligibility and signal-based metric comparison (WER, PESQ, ESTOI) on Voicefixer GSR, Librivox GSR, and CCMusic GSR test sets. WER is not reported for CCMusic GSR due to ASR transcription limitations.}}
    \label{tab:result-wer-pesq}
    \renewcommand{\arraystretch}{1.2}

    \begin{majorrevisiontable}

    \begin{tabular}{llccc}
    \toprule
    \textbf{Dataset} & \textbf{Model} & \textbf{WER$\downarrow$} & \textbf{PESQ$\uparrow$} & \textbf{ESTOI$\uparrow$} \\
    \midrule
    \multirow{6}{*}{Voicefixer GSR}
    & TFGridNet & \textbf{0.092} & \textbf{2.47} & \textbf{0.672} \\
    & NSNet2 & 0.149 & 2.09 & 0.608 \\
    & Voicefixer (checkpoint) & 0.141 & 2.034 & 0.557 \\
    & Voicefixer (retrained) & 0.134 & 2.001 & 0.563 \\
    & MaskSR & 0.147 & 2.319 & 0.658 \\
    \cmidrule{2-5}
    & \tool & 0.113 & 2.449 & 0.667 \\
    \midrule
    \multirow{6}{*}{Librivox GSR}
    & TFGridNet & 0.201 & \textbf{2.076} & 0.708 \\
    & NSNet2 & 0.327 & 1.744 & 0.619 \\
    & Voicefixer (checkpoint) & 0.363 & 1.561 & 0.622 \\
    & Voicefixer (retrained) & 0.24 & 1.588 & 0.663 \\
    & MaskSR & 0.262 & 1.859 & 0.743 \\
    \cmidrule{2-5}
    & \tool & \textbf{0.199} & 2.073 & \textbf{0.784} \\
    \midrule
    \multirow{6}{*}{CCMusic GSR}
    & TFGridNet & - & 1.417 & 0.4 \\
    & NSNet2 & - & 1.474 & 0.396 \\
    & Voicefixer (checkpoint) & - & 1.464 & 0.399 \\
    & Voicefixer (retrained) & - & 1.66 & 0.476 \\
    & MaskSR & - & 2.1 & 0.541 \\
    \cmidrule{2-5}
    & \tool & - & \textbf{2.337} & \textbf{0.585} \\
    \bottomrule
    \end{tabular}

    \end{majorrevisiontable}

\end{table}

\begin{table}[h]
    \centering
    \scriptsize
    \caption{\majorrevision{Evaluation on real-world recordings from the DNS Challenge dataset.}}
    \label{tab:result-dns-real}
    \renewcommand{\arraystretch}{1.2}
    \begin{majorrevisiontable}
    \begin{tabular}{llccc}
    \toprule
    \textbf{Dataset} & \textbf{Model} & \textbf{SIG$\uparrow$} & \textbf{BAK$\uparrow$} & \textbf{OVRL$\uparrow$} \\
    \midrule
    \multirow{6}{*}{\makecell[l]{DNS\\Real Recordings}}
    & TFGridNet & 3.321 & 3.844 & 2.973 \\
    & NSNet2 & 3.037 & 3.963 & 2.76 \\
    & Voicefixer (checkpoint) & 3.289 & 3.955 & 2.988 \\
    & Voicefixer (retrained) & 3.218 & 3.892 & 2.9 \\
    & MaskSR & 3.33 & \textbf{4.044} & 3.056 \\
    \cmidrule{2-5}
    & \tool & \textbf{3.488} & 3.977 & \textbf{3.161} \\
    \bottomrule
    \end{tabular}
    \end{majorrevisiontable}
\end{table}

\majorrevision{We also evaluate the performance of \tool on the DNS Challenge Real Recordings dataset, which contains real-world recordings. The results are shown in Table~\ref{tab:result-dns-real}. \tool achieves the highest SIG and OVRL scores among all models, indicating strong real-world enhancement capability, while maintaining competitive BAK performance.}

\begin{table}[h]
    \centering
    \footnotesize
    \caption{\majorrevision{Training time, model size, and average real-time factor (RTF) comparison on the DNS No Reverb dataset. The RTF is measured on a single GPU. Lower RTF indicates faster inference.}}
    \label{tab:training-rtf}
    \renewcommand{\arraystretch}{1.2}
    \begin{majorrevisiontable}
    \begin{tabular}{lccc}
    \toprule
    \textbf{Model} & \textbf{Model Size} & \textbf{Training Time} & \textbf{RTF$\downarrow$} \\
    \midrule
    Voicefixer (retrained) & 111.7 M & 87.222 h & 0.105 \\
    MaskSR & 361.8 M & 64.118 h & 0.259 \\
    \tool & 363.6 M & 125.833 h & 0.254 \\
    \bottomrule
    \end{tabular}
    \end{majorrevisiontable}
\end{table}

\majorrevision{We report the model size, training time, and average real-time factor (RTF) of Voicefixer, MaskSR, and \tool in Table~\ref{tab:training-rtf}. All models are trained on the same dataset using 4 GPUs, and the RTF is evaluated on the DNS No Reverb dataset using a single GPU. \tool requires the longest training time (125.8 hours) due to its self-critic and REPA design, but achieves competitive inference efficiency (RTF = 0.254) comparable to MaskSR (0.259). While Voicefixer achieves the fastest inference (RTF = 0.105), all models operate well below the real-time threshold (RTF $<$ 1), demonstrating their practical efficiency for deployment.}



\subsubsection{\majorrevision{Effect of Prompt Guidance and Self-Critic Sampling}}

\begin{table*}[htbp]
    \centering
    \footnotesize
    \caption{Effect of Prompt Guidance, \tool can achieve better performance with a speaker prompt.}
    \label{tab:result-ablation-prompt}
    \begin{tabular}{llcccccc}
        \toprule
        \textbf{Dataset} & \textbf{Model} & \textbf{SIG$\uparrow$} & \textbf{BAK$\uparrow$} & \textbf{OVRL$\uparrow$} & \textbf{NISQA$\uparrow$} & \textbf{SpeechBERTScore$\uparrow$} & \textbf{Similarity$\uparrow$} \\
        \midrule
        \multirow{2}{*}{Librivox GSR}
        & \tool & 3.546 & 4.142 & 3.308 & 4.346 & 0.822 & 0.955 \\
        & \tool(w/ prompt) & \textbf{3.636} & \textbf{4.162} & \textbf{3.401} & \textbf{4.472} & \textbf{0.828} & \textbf{0.963} \\
        \midrule
        \multirow{2}{*}{CCMusic GSR}
        & \tool & 3.243 & 3.547 & 2.797 & 3.345 & 0.811 & 0.915 \\
        & \tool(w/ prompt) & \textbf{3.33} & \textbf{3.716} & \textbf{2.944} & \textbf{3.571} & \textbf{0.817} & \textbf{0.921} \\
        \midrule
        \multirow{2}{*}{Voicefixer SR}
        & \tool & 3.449 & \textbf{4.063} & 3.156 & 4.201 & 0.941 & 0.943 \\
        & \tool(w/ prompt) & \textbf{3.477} & 4.041 & \textbf{3.177} & \textbf{4.339} & \textbf{0.946} & \textbf{0.956} \\
        \bottomrule
    \end{tabular}
\end{table*}

\majorrevision{In this section, we investigate the effect of prompt guidance and self-critic sampling on the performance of \tool.} The results in Table~\ref{tab:result-main} were obtained without using any prompt information during inference (except for TSE Group). We first evaluate the performance of \tool with and without prompt guidance on the Librivox GSR, CCMusic GSR, and Voicefixer SR datasets. The results are shown in Table~\ref{tab:result-ablation-prompt}. The model with prompt guidance consistently outperforms the model without prompt guidance across all datasets and metrics (except for BAK on Voicefixer SR). These results indicate that incorporating speaker-specific prompt information during inference can effectively enhance the restoration quality of the model, further demonstrating the adaptability of \tool under speaker-aware scenarios.

\begin{figure*}[htbp]
    \small
    \centering
    \subfloat[Librivox GSR - OVRL\label{fig:ablation-critic-ovrl}]{
        \includegraphics[width=0.32\linewidth]{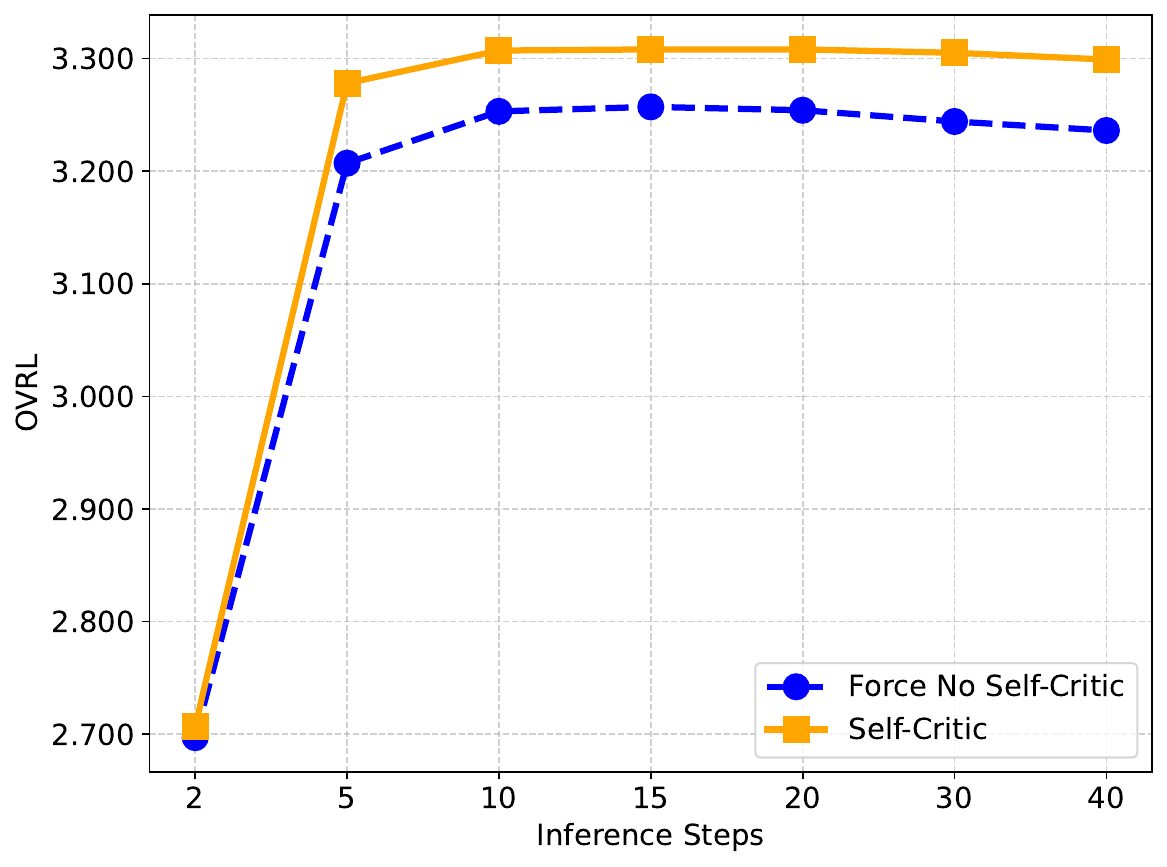}
    }
    \hfill
    \subfloat[Librivox GSR - SpeechBERTScore\label{fig:ablation-critic-sbscore}]{
        \includegraphics[width=0.32\linewidth]{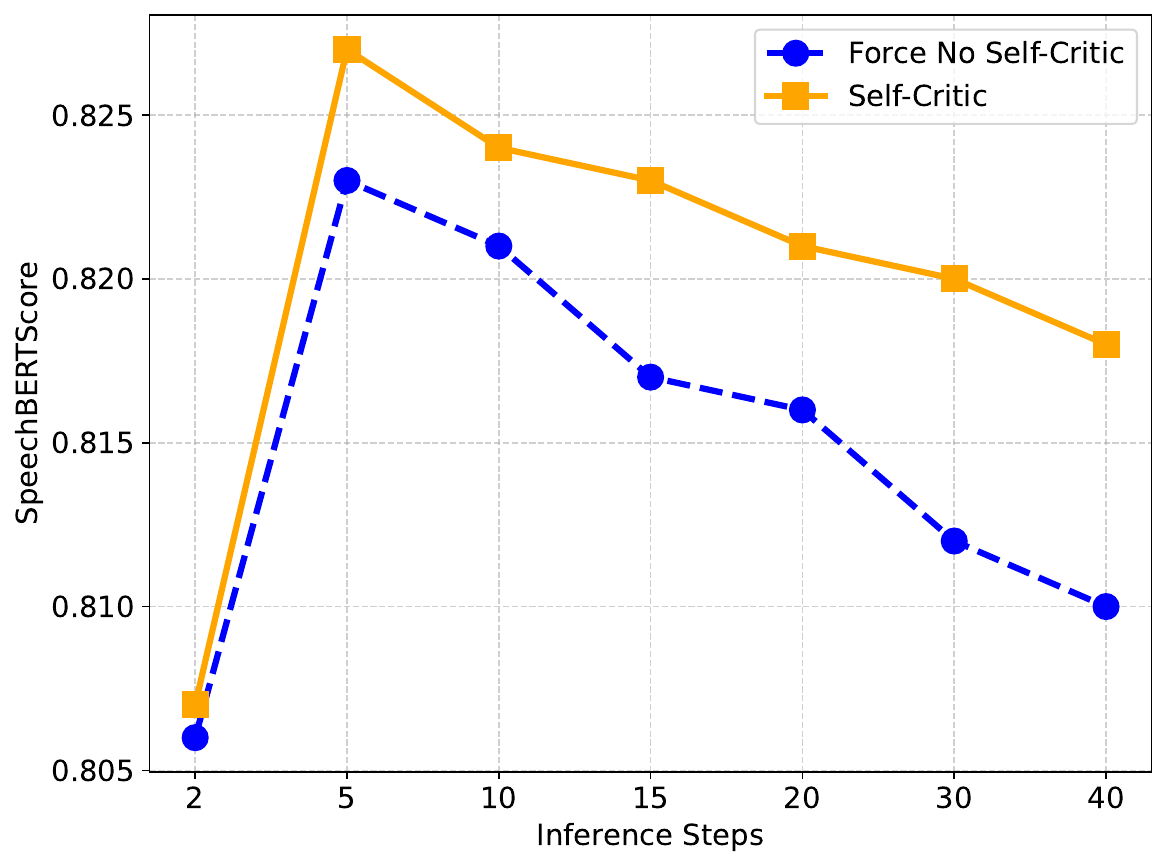}
    }
    \hfill
    \subfloat[Librivox GSR - Similarity\label{fig:ablation-critic-similarity}]{
        \includegraphics[width=0.32\linewidth]{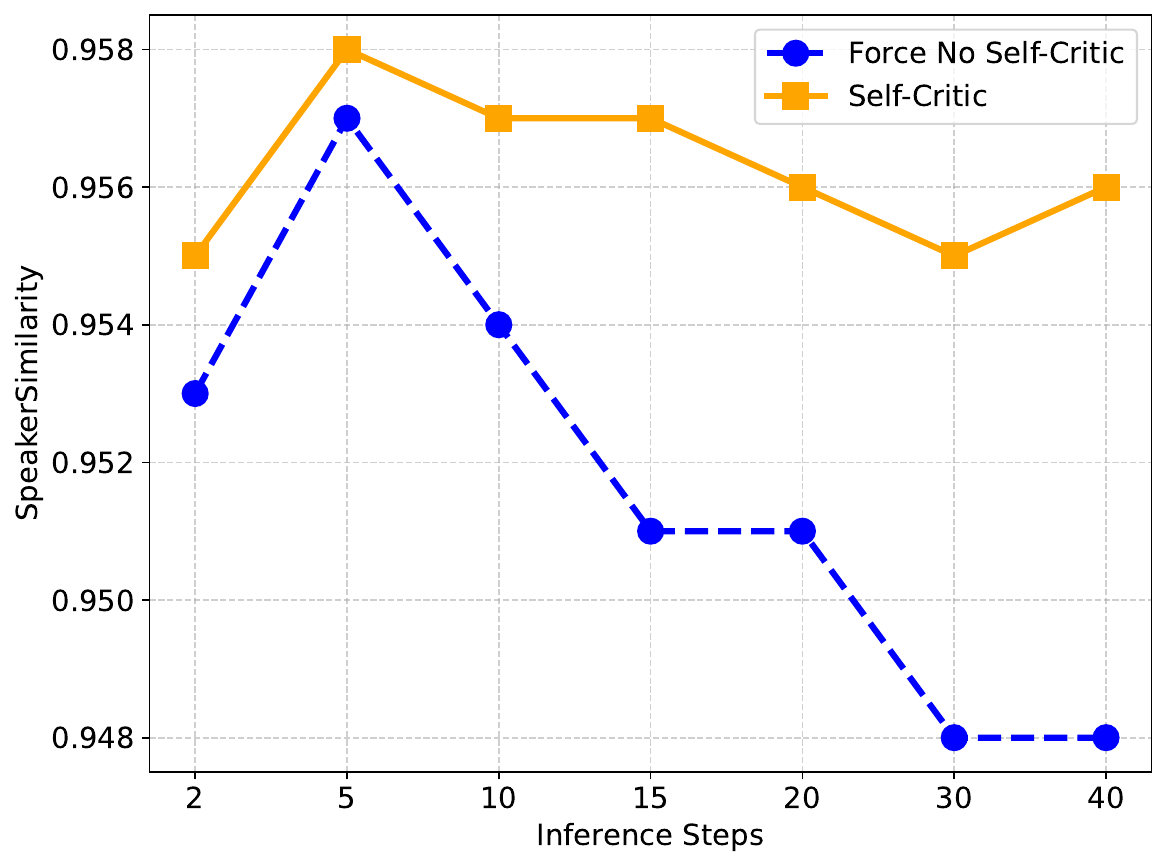}
    }

    \vspace{3pt}

    \subfloat[\majorrevision{CCMusic SR - OVRL}\label{fig:ablation-critic-ccmusic-ovrl}]{
        \includegraphics[width=0.32\linewidth]{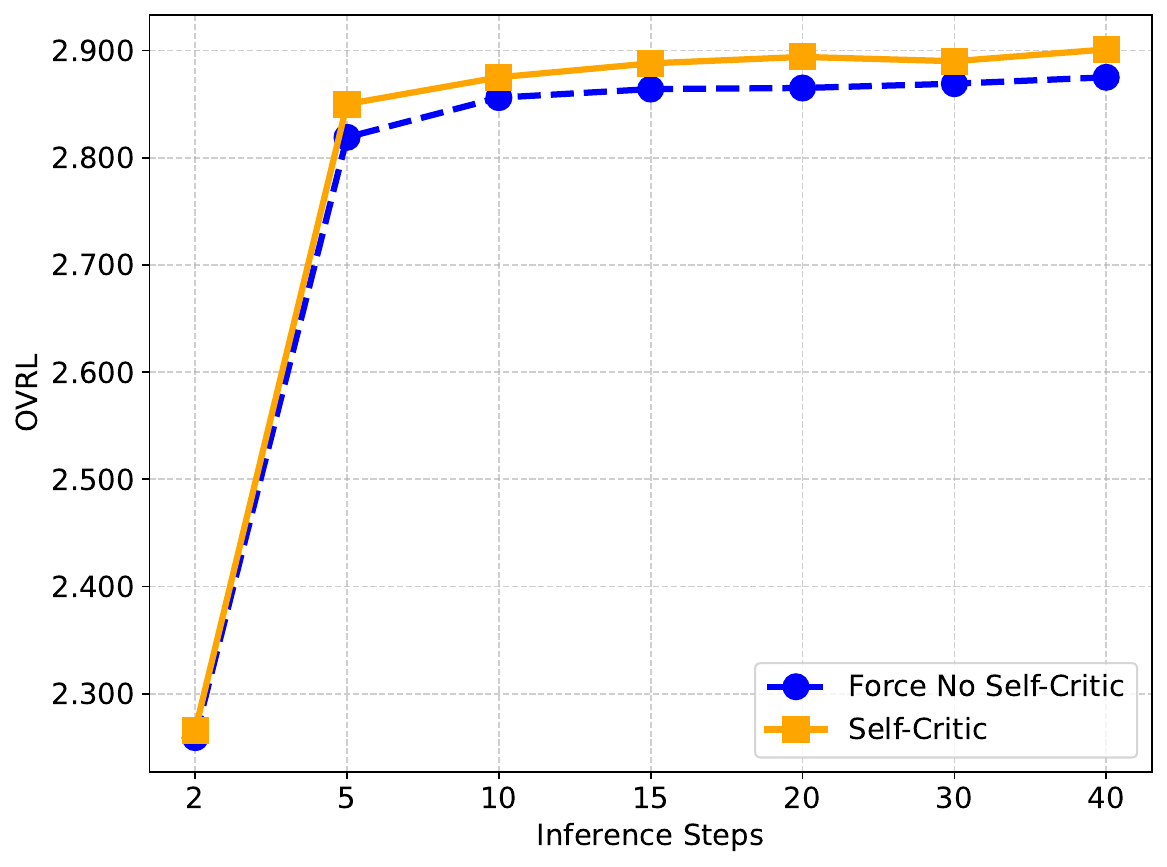}
    }
    \hfill
    \subfloat[\majorrevision{CCMusic SR - SpeechBERTScore}\label{fig:ablation-critic-ccmusic-sbscore}]{
        \includegraphics[width=0.32\linewidth]{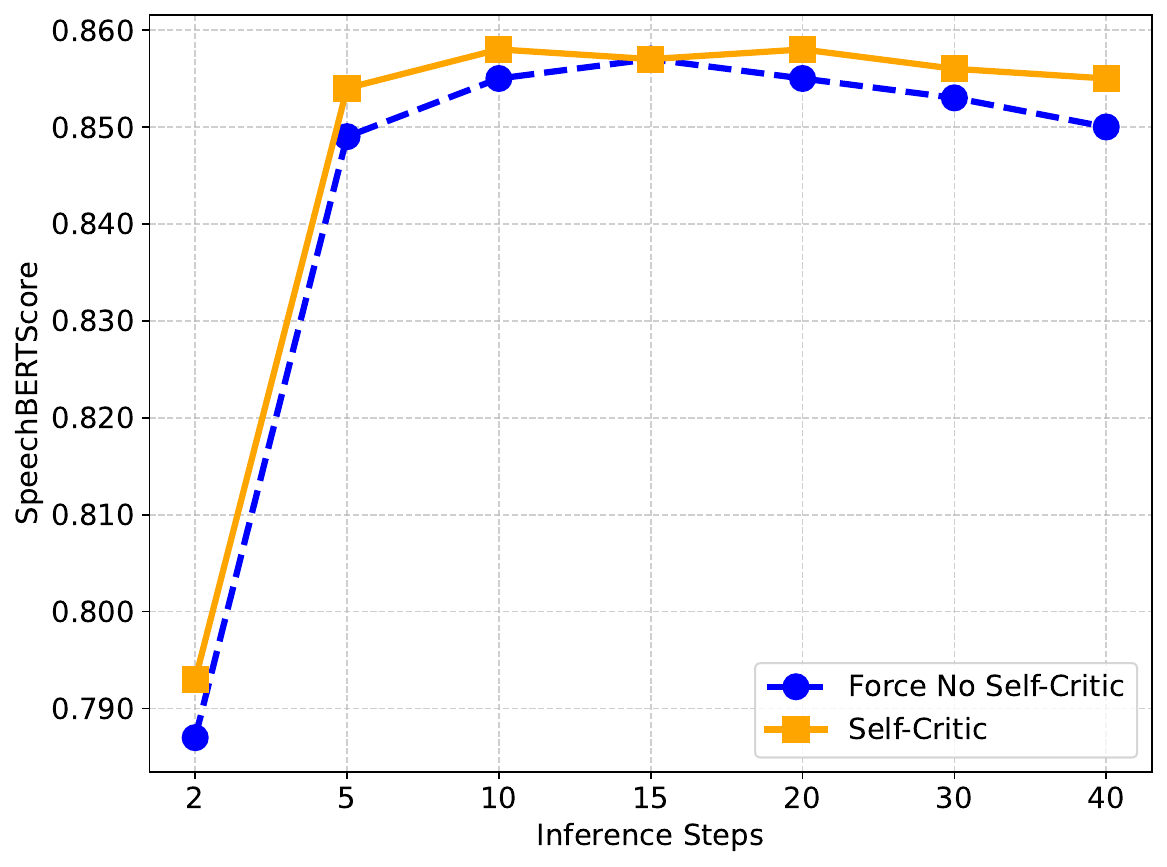}
    }
    \hfill
    \subfloat[\majorrevision{CCMusic SR - Similarity}\label{fig:ablation-critic-ccmusic-similarity}]{
        \includegraphics[width=0.32\linewidth]{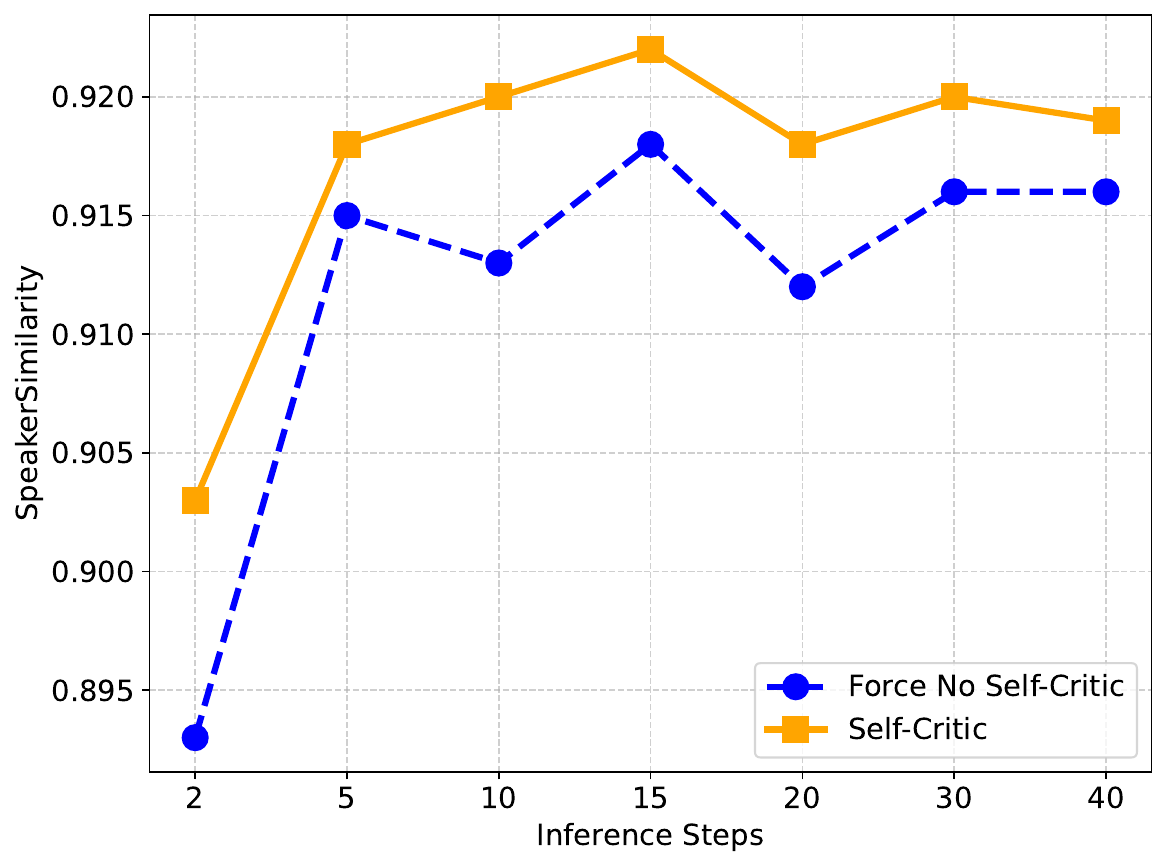}
    }

    \vspace{3pt}

    \subfloat[\majorrevision{VCTK TSE - OVRL}\label{fig:ablation-critic-vctk-ovrl}]{
        \includegraphics[width=0.32\linewidth]{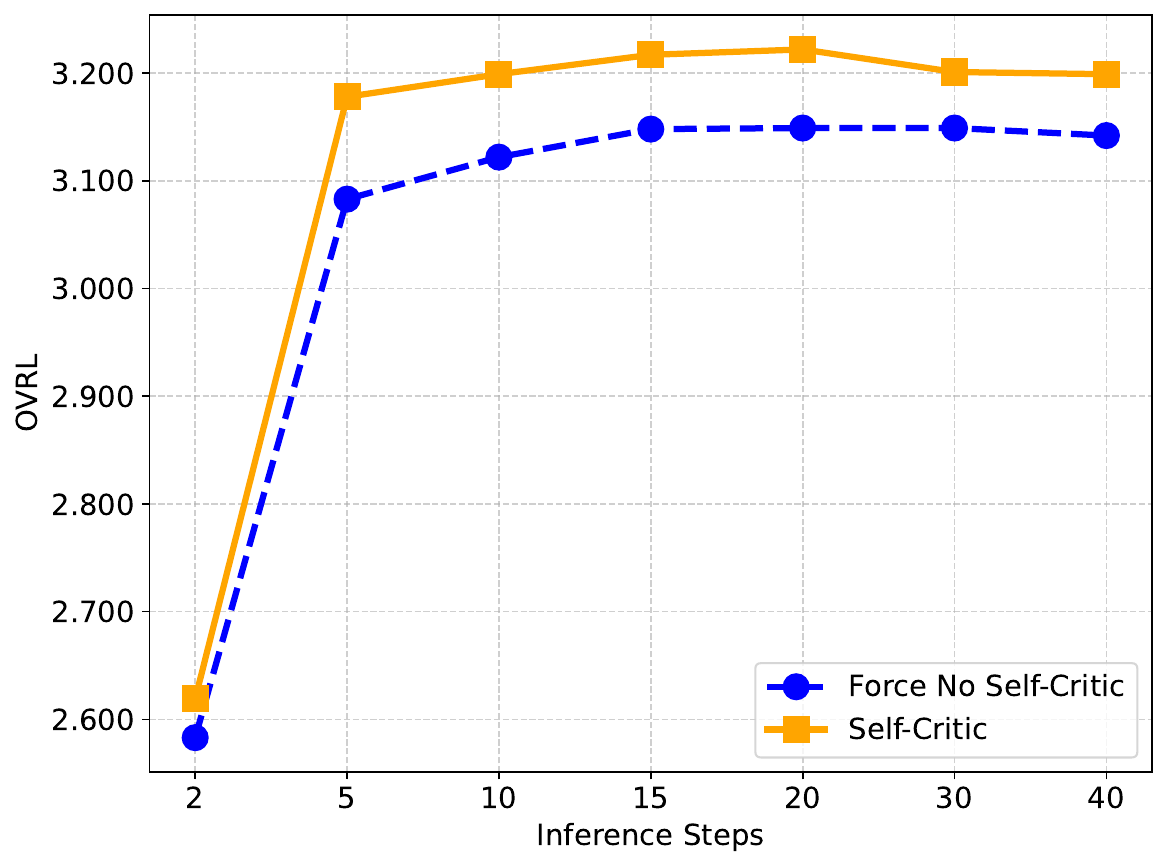}
    }
    \hfill
    \subfloat[\majorrevision{VCTK TSE - SpeechBERTScore}\label{fig:ablation-critic-vctk-sbscore}]{
        \includegraphics[width=0.32\linewidth]{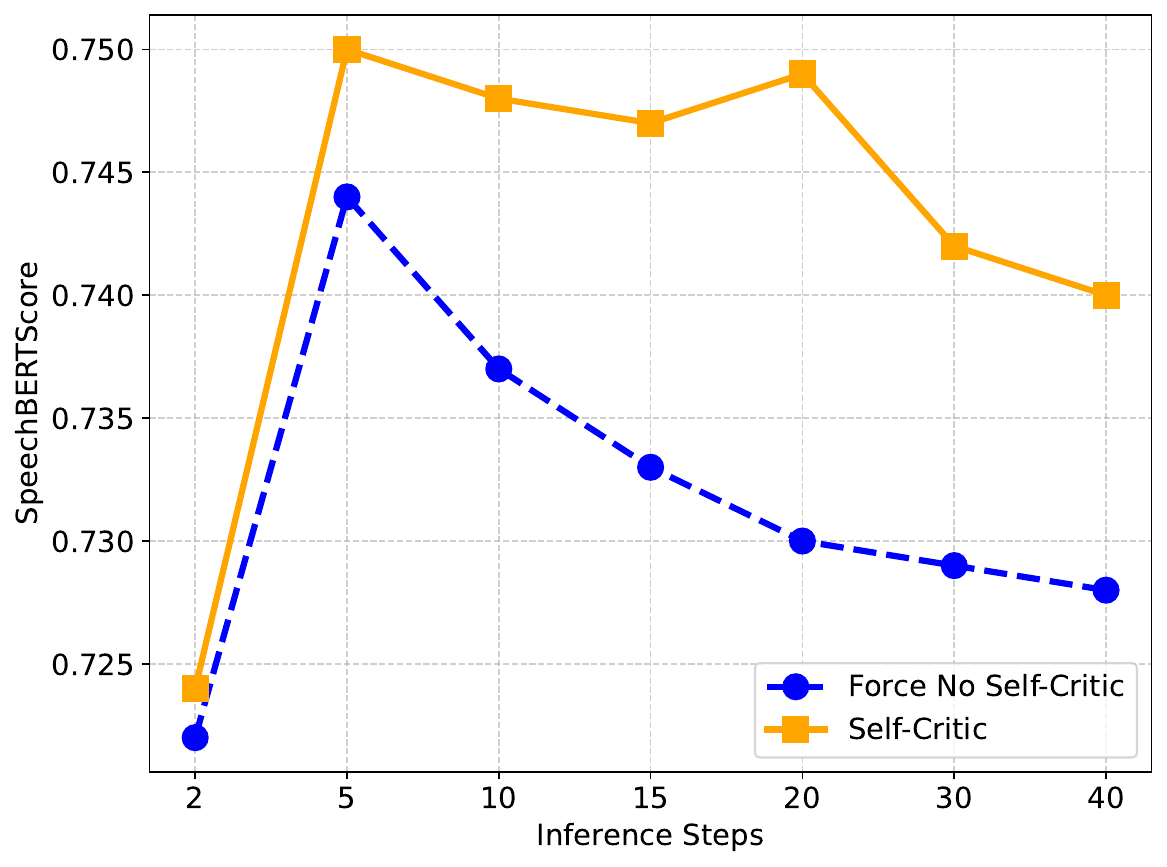}
    }
    \hfill
    \subfloat[\majorrevision{VCTK TSE - Similarity}\label{fig:ablation-critic-vctk-similarity}]{
        \includegraphics[width=0.32\linewidth]{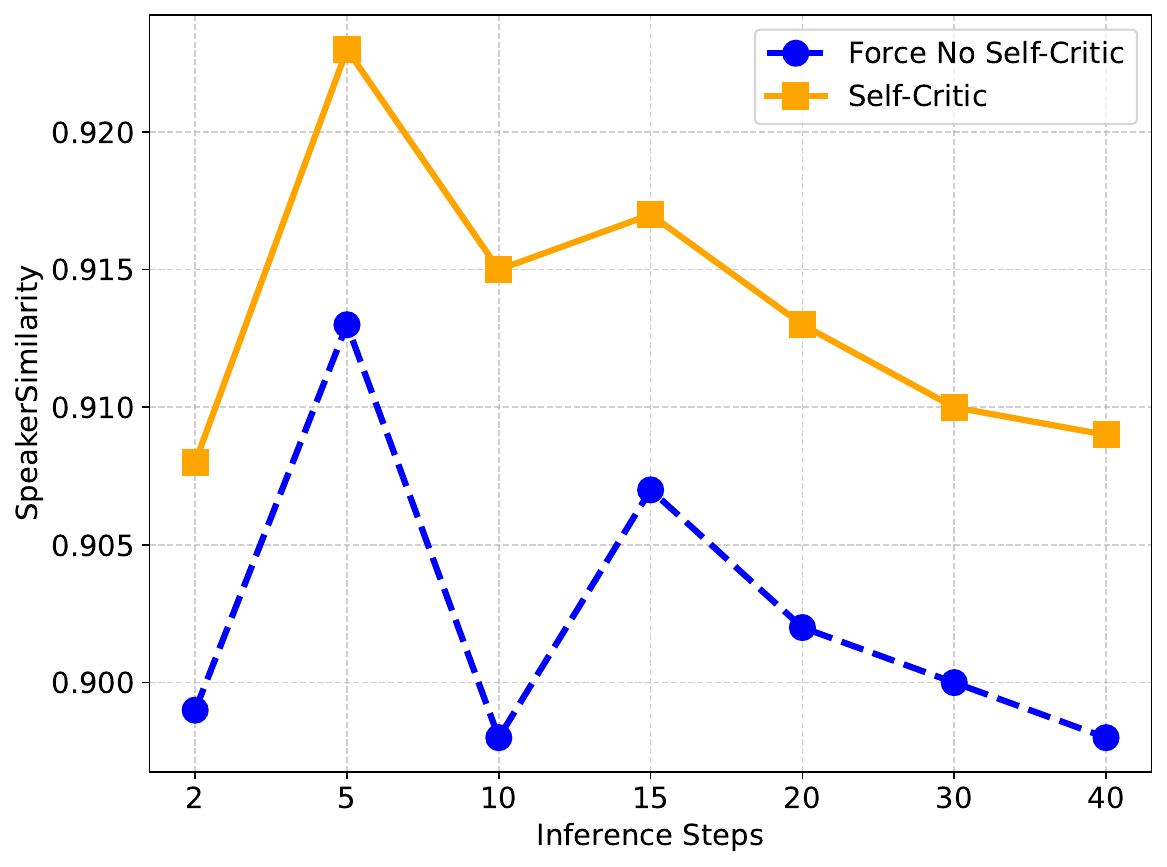}
    }

    \vspace{-4pt}
    \caption{
        Comparison between Self-Critic and non-Self-Critic \majorrevision{across three tasks: Librivox GSR, CCMusic SR, and VCTK TSE.} Using Self-Critic Sampling consistently leads to better performance across varying sampling timesteps and metrics (OVRL, SpeechBERTScore, and Speaker Similarity).
    }
    \label{fig:ablation-critic}
\end{figure*}

\begin{figure*}[htbp]
    \small
    \centering
    \subfloat[\majorrevision{OVRL\label{fig:ablation-critic-prompt-ovrl}}]{
        \includegraphics[width=0.32\linewidth]{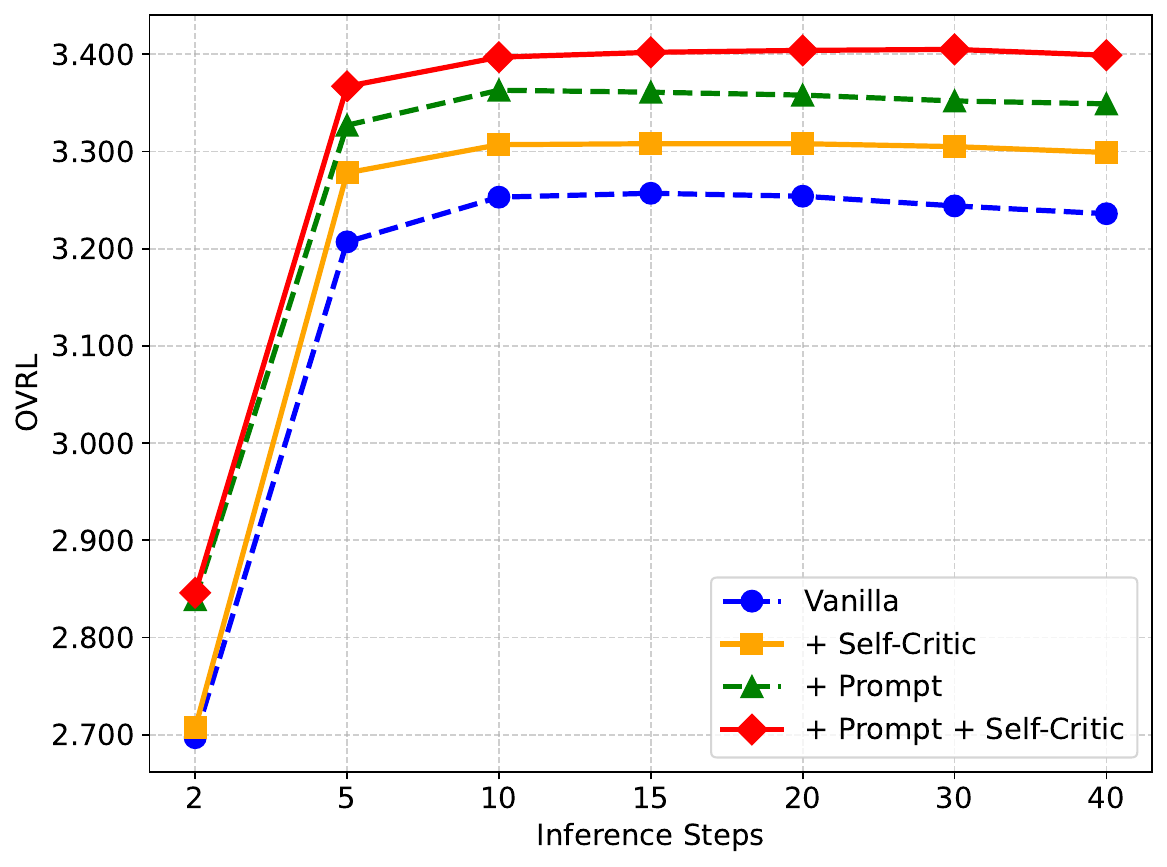}
    }
    \hfill
    \subfloat[\majorrevision{SpeechBERTScore\label{fig:ablation-critic-prompt-sbscore}}]{
        \includegraphics[width=0.32\linewidth]{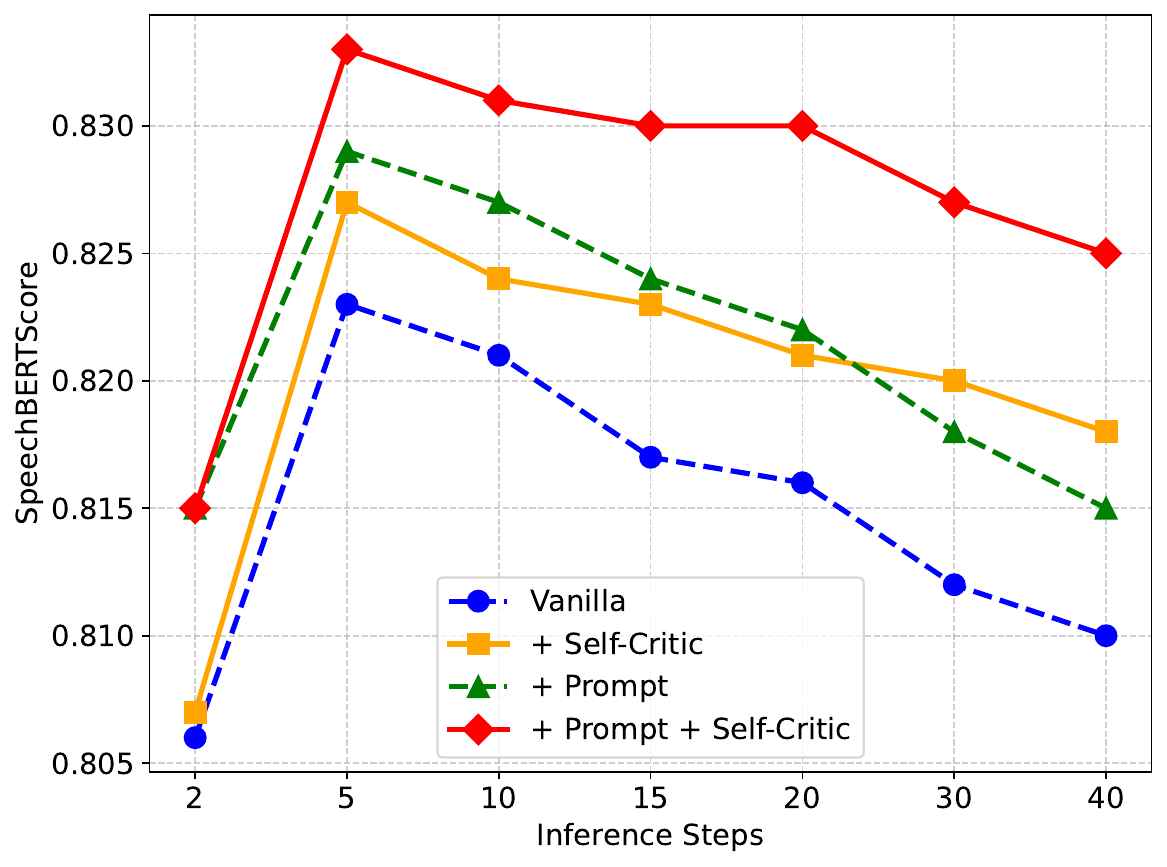}
    }
    \hfill
    \subfloat[\majorrevision{Speaker Similarity\label{fig:ablation-critic-prompt-similarity}}]{
        \includegraphics[width=0.32\linewidth]{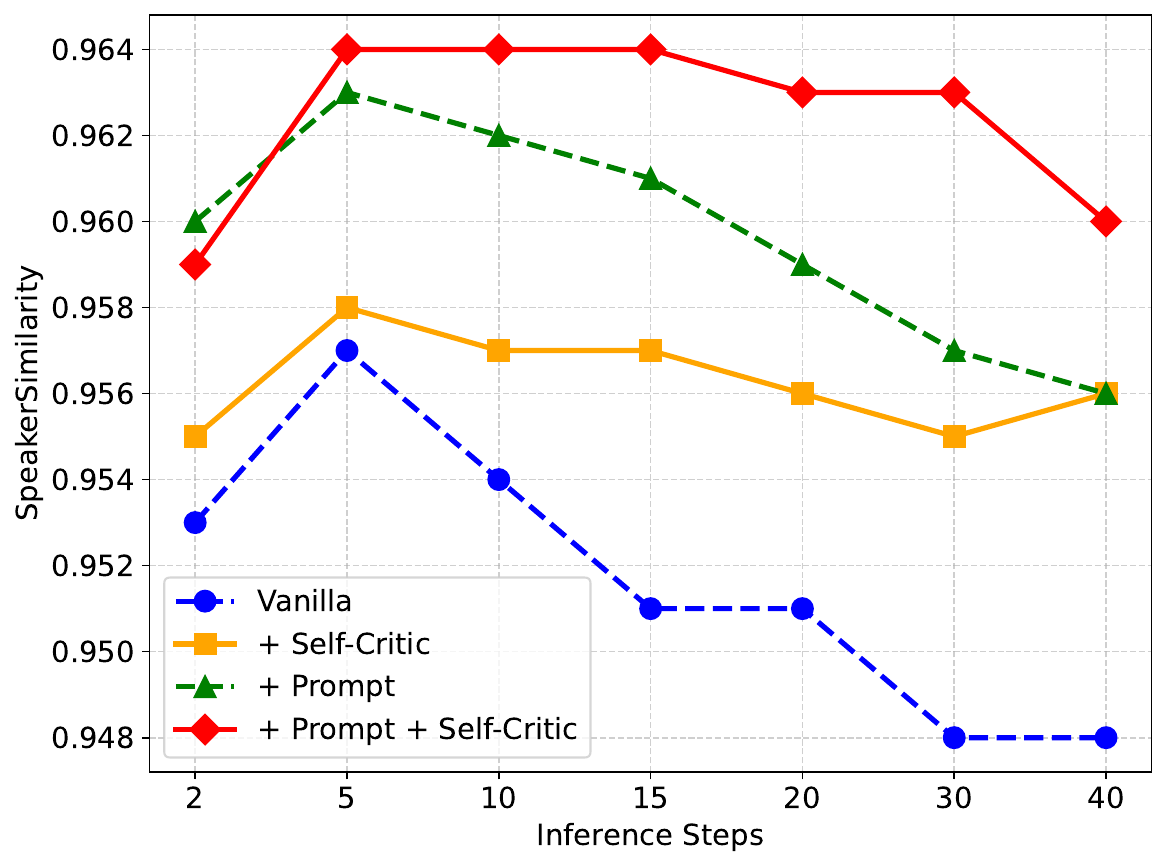}
    }

    \vspace{-4pt}
    \caption{
        \majorrevision{Comparison of the combined and individual effects of Prompt-Guidance and Self-Critic Sampling across different inference steps. Both mechanisms contribute to performance improvements, with prompt-guidance generally offering a larger gain.}
    }
    \label{fig:ablation-critic-prompt}
\end{figure*}

\begin{figure*}[htbp]
    \begin{center}
    \includegraphics[width=0.9\linewidth]{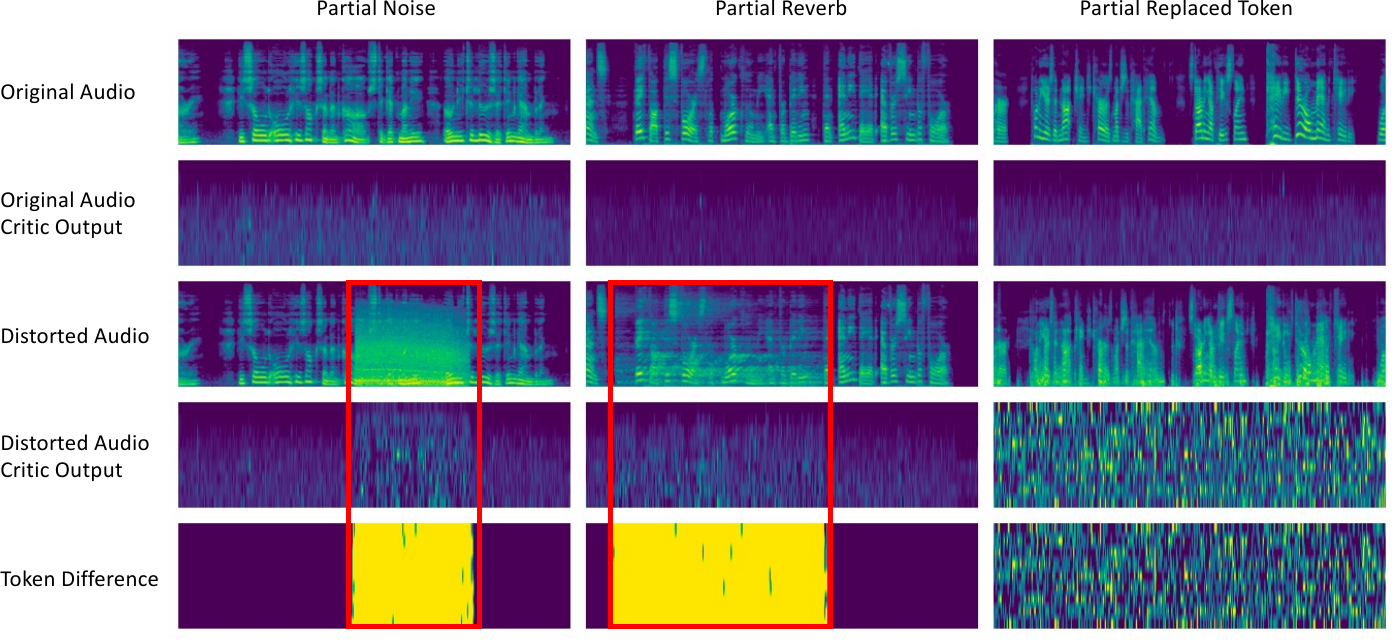}
    \end{center}
    \caption{Visualization of the activations from Critic Head as mentioned in~\ref{sec:self-critic} on partial distorted audio. The critic head as visualized in the fourth row can effectively identify regions affected by noise, reverb, or replaced tokens.}
    \label{fig:critic-head}
\end{figure*}

\majorrevision{To investigate the effectiveness of the self-critic sampling mechanism, we evaluate the model with and without the critic head on 3 different datasets: Librivox GSR, CCMusic SR, and VCTK TSE across different inference steps. The results are shown in Figure~\ref{fig:ablation-critic}.} We observe that the model with the critic head consistently outperforms the model without it across different inference steps, in terms of OVRL, SpeechBERTScore, and Similarity metrics. This demonstrates the importance of the critic head in guiding the model to generate high-quality audio outputs.

To further analyze the role of the critic head, Figure~\ref{fig:critic-head} visualizes the capacity of critic head as mentioned in~\ref{sec:self-critic} to discern distorted audio. We manually constructed cases with partial noise, reverb and replaced acoustic token to the original audio and fed the corresponding audio tokens into the model. The critic head assigns scores to each position, indicating the likelihood of non-realistic audio. The results reveal that the critic head successfully identifies regions affected by these distortions, as evident from the clear distinction in scores between distorted and clean regions. This demonstrates that the critic head effectively learns to differentiate between real and distorted audio, enhancing the model sampling process and improving the quality of the generated audio.

\majorrevision{Finally, we evaluate the combined effect of prompt guidance and self-critic sampling on the model performance. The results are shown in Figure~\ref{fig:ablation-critic-prompt}. We observe that both mechanisms contribute to performance improvements, and prompt guidance generally offering a larger gain. This demonstrates that the two mechanisms can be synergistically combined to enhance the model performance.}

\subsubsection{\majorrevision{Effect of Singing-Voice-oriented Design}}

\majorrevision{In this section, we conduct an ablation study to evaluate the effect of the singing-voice-oriented design on the model performance, namely the improved data simulation process and the inclusion of singing voice data.} First we train the same model on the same dataset with and without the improved data simulation process described in Section~\ref{sec:enhancement_problem}, and evaluate the model on the Librivox GSR and Voicefixer GSR datasets. The Librivox GSR dataset is simulated with the improved data simulation process, while the Voicefixer GSR dataset is simulated using the old-fasioned data simulation process as introduced in Section~\ref{sec:enhancement_problem}. The results are shown in Table~\ref{tab:result-ablation-data-simulation}. Even though the Voicefixer GSR dataset is not simulated with the improved data simulation process, the model trained with the improved data simulation process still outperforms the model trained without it across all metrics (except BAK on the Voicefixer GSR dataset), demonstrating the effectiveness of the improved data simulation process in enhancing generalization and robustness of \tool.

\majorrevision{To better understand the singing-voice-oriented design, we conduct an ablation study to evaluate both the improved data simulation process and the inclusion of singing voice data on the model performance. We train the model with and without the improved data simulation process and the inclusion of fullband singing voice dataset. The results are shown in Table~\ref{tab:result-ablation-singing}. Both the improved data simulation process and the inclusion of singing voice data contribute to the performance improvements across all metrics, with the improved data simulation process introduces gain on the overall quality (OVRL, NISQA), and the inclusion of singing voice data improves largely on the performance on SpeechBERTScore and speaker similarity. The model trained with both enhancements achieves the best performance, demonstrating the effectiveness of these design choices in enhancing the model's ability to handle singing voice scenarios.}

\begin{table*}[htbp]
    \centering
    \footnotesize
    \caption{Effect of Improved Data Simulation Process, w/o improved data simulation means the model is trained using $ d_{\text{noise}}(d_{\text{bw}}(d_{\text{clip}}(d_{\text{reverb}}(\cdot)))) $ described in Section~\ref{sec:enhancement_problem}, w/ improved data simulation means the model is trained using $ d_{\text{bw}}(d_{\text{clip}}(d_{\text{reverb}}(d_{\text{noise}}(d_{\text{vocal\_effect}}(\cdot))))) $ in addition to traditional data simulation.}
    \label{tab:result-ablation-data-simulation}
    \begin{tabular}{llcccccc}
        \toprule
        \textbf{Dataset} & \textbf{Model} & \textbf{SIG$\uparrow$} & \textbf{BAK$\uparrow$} & \textbf{OVRL$\uparrow$} & \textbf{NISQA$\uparrow$} & \textbf{SpeechBERTScore$\uparrow$} & \textbf{Similarity$\uparrow$} \\
        \midrule
        \multirow{2}{*}{Librivox GSR}
        & \tool(w/o improved data simulation) & 3.361 & 4.096 & 3.093 & 3.766 & 0.731 & 0.888 \\
        & \tool(w/ improved data simulation) & \textbf{3.504} & \textbf{4.135} & \textbf{3.255} & \textbf{4.191} & \textbf{0.762} & \textbf{0.91} \\
        \midrule
        \multirow{2}{*}{Voicefixer GSR}
        & \tool(w/o improved data simulation) & 3.355 & \textbf{4.033} & 3.066 & 4.145 & 0.798 & 0.897 \\
        & \tool(w/ improved data simulation) & \textbf{3.406} & 4.024 & \textbf{3.11} & \textbf{4.339} & \textbf{0.832} & \textbf{0.909} \\
        \bottomrule
    \end{tabular}
\end{table*}

\begin{table*}[htbp]
    \centering
    \footnotesize
    \caption{\majorrevision{Ablation study on singing-voice-oriented design. We examine the effects of (1) improving the data simulation process and (2) including singing voice data. Results show improvements across all evaluation metrics, particularly in SpeechBERTScore and speaker similarity.}}
    \label{tab:result-ablation-singing}
    \begin{majorrevisiontable}

    \begin{tabular}{lcccccccc}
        \toprule
        \textbf{Dataset} & \textbf{+ Improved Simulation} & \textbf{+ Singing Voice Data} & \textbf{SIG$\uparrow$} & \textbf{BAK$\uparrow$} & \textbf{OVRL$\uparrow$} & \textbf{NISQA$\uparrow$} & \textbf{SpeechBERTScore$\uparrow$} & \textbf{Similarity$\uparrow$} \\
        \midrule
        \multirow{4}{*}{CCMusic GSR}
        & \ding{55} & \ding{55} & 2.981 & 3.736 & 2.542 & 2.916 & 0.559 & 0.776 \\
        & \ding{51} & \ding{55} & 3.17 & \textbf{3.77} & 2.75 & \textbf{3.406} & 0.598 & 0.839 \\
        & \ding{55} & \ding{51} & 3.144 & 3.537 & 2.694 & 3.124 & 0.739 & 0.863 \\
        & \ding{51} & \ding{51} & \textbf{3.213} & 3.647 & \textbf{2.809} & 3.358 & \textbf{0.759} & \textbf{0.889} \\
        \bottomrule
    \end{tabular}

    \end{majorrevisiontable}

\end{table*}

\subsubsection{\majorrevision{Effect of REPA module and Semantic Feature}}

\begin{table*}[htbp]
    \centering
    \footnotesize
    \caption{\majorrevision{Ablation study on REPA module and semantic features across Librivox GSR and Voicefixer GSR test sets. Both REPA and semantic features contribute to performance improvements.}}
    \label{tab:result-ablation-feature-combined}
    \vspace{-5pt}
    \begin{majorrevisiontable}
    \begin{tabular}{llcccccc}
        \toprule
        \textbf{Dataset} & \textbf{Model} & \textbf{SIG$\uparrow$} & \textbf{BAK$\uparrow$} & \textbf{OVRL$\uparrow$} & \textbf{NISQA$\uparrow$} & \textbf{SpeechBERTScore$\uparrow$} & \textbf{Similarity$\uparrow$} \\
        \midrule
        \multirow{3}{*}{Librivox GSR}
        & \tool & \textbf{3.505} & 4.135 & \textbf{3.256} & \textbf{4.200} & 0.763 & 0.913 \\
        & \quad - REPA & 3.481 & \textbf{4.141} & 3.242 & 4.082 & \textbf{0.764} & \textbf{0.923} \\
        & \quad \quad - Noisy Emb & 3.430 & 4.123 & 3.178 & 3.975 & 0.734 & 0.902 \\
        \midrule
        \multirow{3}{*}{Voicefixer GSR}
        & \tool & \textbf{3.400} & 4.022 & \textbf{3.103} & \textbf{4.329} & \textbf{0.831} & 0.907 \\
        & \quad - REPA & 3.372 & \textbf{4.024} & 3.082 & 4.210 & 0.830 & \textbf{0.915} \\
        & \quad \quad - Noisy Emb & 3.343 & 4.051 & 3.061 & 4.215 & 0.812 & 0.905 \\
        \bottomrule
    \end{tabular}
    \end{majorrevisiontable}
\end{table*}

\majorrevision{We examine the effect of the REPA module and semantic features on both the Librivox GSR and Voicefixer GSR test sets, as shown in Table~\ref{tab:result-ablation-feature-combined}. Removing the REPA module leads to slight degradations in SIG, OVRL, and NISQA, indicating that REPA contributes to perceptual and quality improvements. Further removing the noisy semantic features results in consistent drops across all metrics, especially SpeechBERTScore and Similarity, highlighting the effectiveness of the semantic features in enhancing the model's ability.}

\subsubsection{Effect of One Model for All Tasks}

\begin{table*}[htbp]
    \centering
    \footnotesize
    \caption{Effect of one model for all tasks, we compare models trained on each single task's distortion against a unified model trained on all tasks simultaneously. The unified multitask training approach consistently outperforms single-task model across all datasets.}
    \label{tab:result-ablation-task}
    \begin{tabular}{llcccccc}
        \toprule
        \textbf{Dataset} & \textbf{Model} & \textbf{SIG$\uparrow$} & \textbf{BAK$\uparrow$} & \textbf{OVRL$\uparrow$} & \textbf{NISQA$\uparrow$} & \textbf{SpeechBERTScore$\uparrow$} & \textbf{Similarity$\uparrow$} \\
        \midrule
        \multirow{2}{*}{Librivox GSR}
        & \tool(GSR) & 3.438 & 4.112 & 3.179 & 3.827 & 0.74 & \textbf{0.913} \\
        & \tool & \textbf{3.504} & \textbf{4.135} & \textbf{3.255} & \textbf{4.191} & \textbf{0.762} & 0.91 \\
        \midrule
        \multirow{2}{*}{DNS No Reverb$^{*}$}
        & \tool(SE) & \textbf{3.646} & 4.189 & 3.425 & 4.765 & \textbf{0.877} & \textbf{0.981} \\
        & \tool & 3.643 & \textbf{4.205} & \textbf{3.432} & \textbf{4.774} & 0.876 & 0.98 \\
        \midrule
        \multirow{2}{*}{Voicefixer SR}
        & \tool(SR) & 3.443 & 4.04 & 3.141 & 4.144 & \textbf{0.943} & \textbf{0.941} \\
        & \tool & \textbf{3.476} & \textbf{4.085} & \textbf{3.197} & \textbf{4.339} & 0.923 & 0.926 \\
        \midrule
        \multirow{2}{*}{VCTK TSE}
        & \tool(TSE) & 3.479 & 4.004 & 3.17 & 4.251 & 0.596 & 0.888 \\
        & \tool & \textbf{3.465} & \textbf{4.077} & \textbf{3.185} & \textbf{4.451} & \textbf{0.653} & \textbf{0.9} \\
        \bottomrule
    \end{tabular}
\end{table*}

In this ablation study, we compare the performance of \tool trained individually on each task against a unified model trained on all tasks simultaneously. The results are shown in Table~\ref{tab:result-ablation-task}. The results show that the unified model outperforms task-specific models across most metrics. For instance, on the Librivox GSR dataset, the unified model achieves higher SIG (3.504 vs. 3.438) and NISQA (4.191 vs. 3.827) scores compared to the task-specific model. Similar trends are observed in other tasks, such as SE, SR, and TSE, where the unified model achieves consistently higher OVRL and NISQA scores.

We hypothesize that training on all tasks together introduces diversity to the input data, effectively increasing the complexity of the distortion function $D(\cdot)$ that the model must learn to invert (See Section~\ref{sec:enhancement_problem}). By jointly learning to recover clean audio across diverse scenarios, the unified model better approximates the inverse function $D^{-1}(\cdot)$, resulting in improved overall performance. This can be likened to a form of data augmentation, where the presence of multiple distortion types and tasks enriches the training distribution. Consequently, the model becomes more robust to various distortions and achieves better generalization.

%% file: src/ch05-conclusion.tex
\section{Conclusion}

In this work, we introduce \tool, a unified voice enhancement model for both speech and singing voice. Using a masked generative model as the base, \tool is capable of addressing diverse enhancement tasks, including speech denoising, dereverberation, super-resolution, and target speaker extraction. We incorporate a prompt-guidance mechanism for in-context learning, which allows the model to natively accept a reference speaker's timbre without altering the underlying architecture. This strategy boosts enhancement performance when a reference audio is available and enable the target speaker extraction task without altering the underlying architecture. Moreover, we integrate a self-critic mechanism into the generative process for masked generative models, yielding higher-quality outputs through iterative self-assessment and refinement. Extensive experiments on various enhancement tasks demonstrate that \tool outperforms existing methods in terms of both objective metrics and subjective listening tests. \majorrevision{Future work includes constructing more realistic degradation simulation scenarios, exploring textual prompts for guided enhancement, and learning optimal task-specific loss weights for multitask training.}